# Wireless Energy Transfer for Space Applications: A Review of Technologies, Mission Architectures, Energy Routing, and System Integration

Lyuye Lin[1,2], Zhifu Feng[2], Ermanno Miele[3], Quanbo Ge[1], Manuel Iori[4], Chunbo Liu[1], Kwok-Tong Chau[6], Denis Garoli[2,4*] and Remo Proietti Zaccaria[2*]

[1] School of Electronic and Information Engineering, Nanjing University of Information Science and Technology, Nanjing, China

[2] Istituto Italiano di Tecnologia, Via Morego 30, 16163 Genova (Italy)

[3] University of Cambridge, Cavendish Laboratory, Cambridge (UK)

[4] Università degli Studi di Modena e Reggio Emilia, Modena (Italy)

[5] Department. of Mechanical and Electrical Engineering, Henan University of Technology, China (CN)

[6] Department of Electrical and Electronic Engineering, The Hong-Kong Polytechnique University (HK)

Corresponding authors: Prof. Garoli Denis – denis.garoli@unimore.it; Dr. Remo Proietti Zaccaria – remo.proietti@iit.it

## ABSTRACT

Wireless energy transfer (WET) decouples the location of energy generation from that of energy use, a capability of growing strategic importance as space missions evolve toward distributed spacecraft architectures, lunar surface networks, mobile assets, and high-power remote loads. This review organizes the field along four axes: physical mechanism, energy reference plane, mission architecture, and evidence maturity. Near-field inductive and resonant transfer, microwave and millimeter-wave power beaming, optical and laser-based transfer, space-based solar power, lunar and cislunar energy delivery, inter-spacecraft power sharing, intra-spacecraft contactless interfaces, and reconfigurable or relay-assisted routing are each assessed within a common R0–R7 energy-accounting framework. Three overarching conclusions emerge. First, WET is most compelling when a mission presents a genuine spatial or temporal mismatch between energy generation and load demand; it is not a universal substitute for local generation or onboard energy storage. Second, dominant loss mechanisms shift across the system architecture, so metrics such as beam-collection efficiency, rectification efficiency, receiver electrical output, and useful-load efficiency cannot be meaningfully compared without precisely defined energy reference planes. Third, experimental maturity remains highly uneven: high-power ground demonstrations and short-range flight applications are relatively well established, whereas long-range radiative inter-spacecraft power transfer, spaceborne routing hardware, lunar energy networks, and deep-space applications remain largely supported by models, simulations, and mission studies rather than hardware demonstration. Closing these gaps will require validated, end-to-end energy budgets under mission-realistic constraints, encompassing pointing and tracking, thermal management, environmental exposure, reliability, energy storage, and regulatory compliance.

## 1. Introduction

Electrical power is increasingly an architectural resource rather than a self-contained spacecraft subsystem. Conventional spacecrafts are designed around local energy generation and storage: photovoltaic arrays exploit available illumination, batteries bridge eclipse and transient demand, and nuclear sources serve missions for which solar energy is weak, intermittent, or operationally impractical. Distributed exploration increasingly challenges this paradigm. Permanently shadowed lunar assets can be separated by kilometers from illuminated terrain; mobile rovers can move beyond fixed power infrastructure; formation-flying spacecraft can encounter different illumination and load histories; and planetary or deep-space clients may operate where the most favorable generation site is an orbiter, surface, aerial, or mothercraft platform [1–4]. The common systems problem is therefore not simply insufficient power, but a spatial and temporal mismatch between where energy can be generated economically and where and when mission functions require it. Throughout this Review, local generation, storage, and wired distribution are treated as reference comparators for WET. WET is therefore evaluated architecturally rather than at the component level, including transmitter, receiver, pointing, thermal, control, and residual-storage burdens. WET introduces a new degree of freedom by allowing generation, transmission, reception and consumption to occupy different physical locations. Depending on scale, this can be achieved through inductive or resonant near-field coupling, radiative microwave or millimeter-wave transmission, or optical/laser power beaming [5–7]. At short range, contactless interfaces can reduce dependence on sliding contacts, penetrations, and connectors. At longer range, WET supports spatial energy delivery from orbit to Earth, from illuminated lunar or orbital nodes to shadowed loads or between spacecrafts. Reconfigurable apertures and relays can additionally alter energy paths when line of sight, pointing range or multi-user service becomes limiting [8–10]. The literature has developed along several strong but only partially connected traditions. Space-based solar power (SBSP) established an early and extensively studied mission-scale lineage, from early solar power satellite concepts to high-power terrestrial tests and recent coherent-array flight experiments [11–13]. The broader wireless power transfer (WPT) literature established the underlying coupling, conversion, rectification, and waveform-design principles [14–16]. Non-terrestrial-network research now treats energy as a schedulable resource across multiple nodes [17], while safety and exposure studies have developed largely in parallel [18–21]. Recent SBSP reviews and system-level studies provide valuable perspectives on architecture, dispatchability, and technoeconomic feasibility [22, 23]. Yet cross-comparison remains difficult because efficiencies are often reported at different electrical or electromagnetic boundaries, evidence classes are mixed, and application studies frequently optimize the link without closing the source, storage, auxiliary-power or useful-load budgets. To enable consistent comparison, reported results are interpreted throughout this Review according to their evidence class, distinguishing peer-reviewed experimental demonstrations, agency-reported tests, simulations and mission studies, and technoeconomic projections.

In this paper we examine the mission conditions under which wireless transfer can move useful energy more effectively than the best conventional alternative. The discussion spans near-field contactless transfer; microwave, millimeter-wave and optical power beaming; Space-to-Earth SBSP; lunar and cislunar delivery; spacecraft-to-spacecraft sharing; selected deep-space applications; intra-spacecraft interfaces; reconfigurable apertures and relays; enabling RF and optical hardware; environmental qualification; safety; reliability; and the transition toward distributed energy service. Beamed propulsion, including solar-sail and laser-driven lightsail concepts, is outside the scope of this review because the primary delivery quantity is photon momentum for propulsion rather than electrical energy at a remote load [24–26]. Communication-only reconfigurable-surface and simultaneous information-and-power-transfer studies are considered only where they directly inform an energy-transfer architecture. Accordingly, the contribution is not a new universal efficiency metric, but a common framework for determining what each experiment or mission study actually demonstrates and what remains unclosed.

## 2. Fundamentals and Evolution of Wireless Energy Transfer for Space Systems

WET delivers electrical energy through electromagnetic coupling or radiation, with the dominant transfer mechanism determined by the electrical dimensions, separation, and operating conditions of the source–receiver system. Reactive near-field transfer relies primarily on magnetic or electric coupling, whereas microwave, millimeter-wave, and optical transfer require electromagnetic field generation, propagation, reception, and electrical conversion [5, 6, 14, 27]. These regimes cannot be separated by universal distance thresholds because wavelength, aperture dimensions, and link geometry jointly determine the applicable field description. Figure 1 connects representative space applications with the corresponding energy-transfer pathways and their principal design considerations. Near-field systems are governed mainly by coupling, resonance, impedance matching, and alignment. Radiative microwave and millimeter-wave systems additionally depend on beam formation, finite-aperture interception, and RF-to-DC conversion, while optical systems place greater emphasis on beam quality, pointing, wavelength matching, and photovoltaic thermal management. These differences explain why an architecture suitable for a contactless spacecraft interface cannot be assessed using the same design assumptions as a long-range power-beaming link. This section develops the physical mechanisms governing WET and their implications for space systems. Figure 2 provides an overview of electromagnetic energy transport, coupling, propagation, reception, and conversion, together with the processes that limit useful power delivery. Section 2.1 establishes their common physical basis, followed by mechanism-specific discussions in Sections 2.2–2.5. Section 3 subsequently introduces the R0–R7 reference-plane framework for evaluating complete architectures using consistently defined power and energy boundaries.

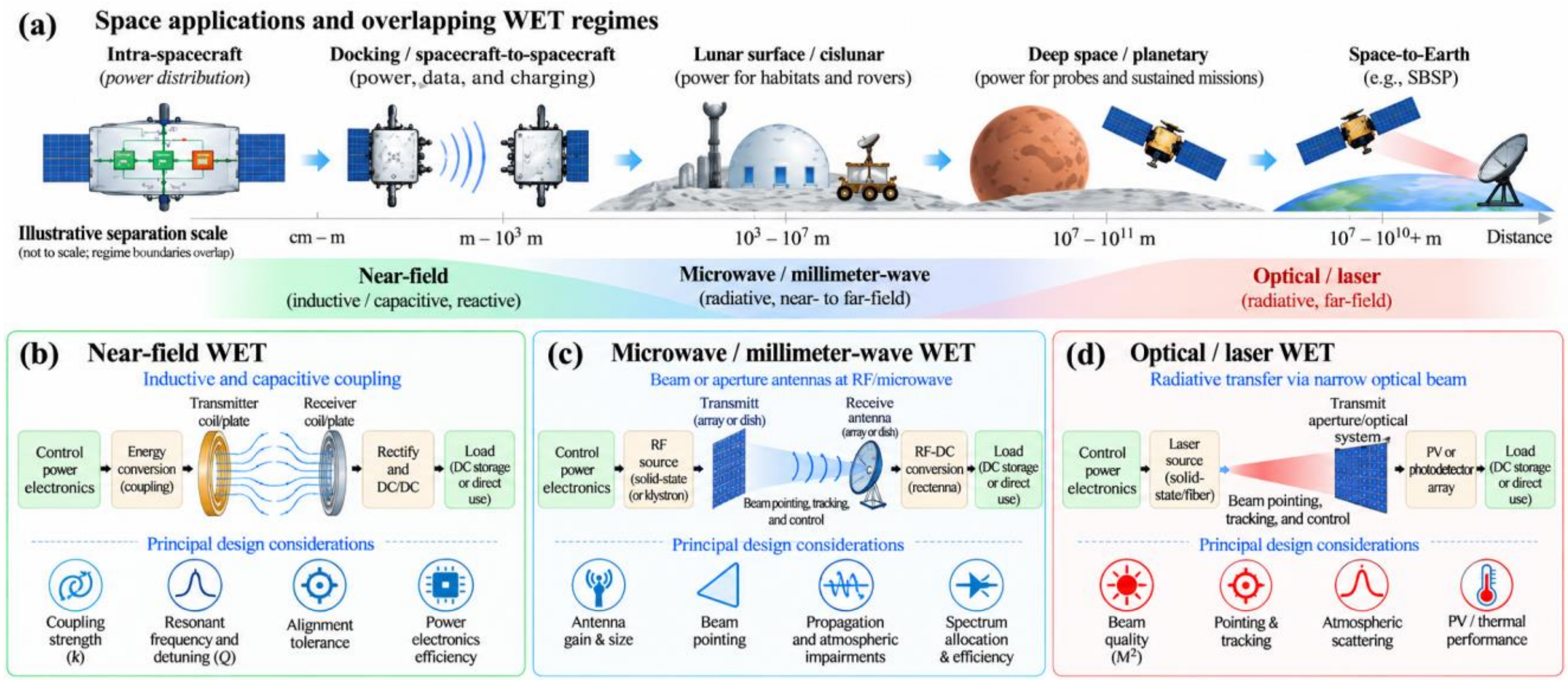

**Figure 1. Space wireless energy transfer across mission scales and transmission regimes.** (a) Illustrative space applications and their characteristic separation scales, from intra-spacecraft distribution to space-to-Earth power beaming. The overlapping bands indicate applicable WET regimes rather than fixed distance boundaries. (b) Near-field inductive and capacitive transfer, highlighting coupling strength, resonance and impedance matching, and alignment tolerance. (c) Microwave and millimeter-wave transfer, showing RF generation, beam formation, propagation, reception, and rectification, together with the effects of aperture size, pointing, propagation loss, and conversion efficiency. (d) Optical transfer, illustrating laser generation, beam propagation, photovoltaic reception, and power conditioning, with beam quality, pointing, spectral matching, and thermal management as principal design considerations. The schematic pathways do not share a common efficiency reference plane.

## 2.1 Physical Basis of WET: From Electromagnetic Field to Useful Power

WET is governed by electromagnetic energy conservation, irrespective of the source, receiver, or transmission architecture. For time-harmonic fields represented by peak-amplitude phasors, the time-averaged Poynting vector gives the net active power flux:

$$\langle \boldsymbol{S} \rangle = \frac{1}{2} Re(\boldsymbol{E} \times \boldsymbol{H}^*) \tag{1}$$

This expression follows from the time-harmonic formulation of Poynting's theorem [28, 29]. Here, $\boldsymbol{E}$ and $\boldsymbol{H}$ denote the electric- and magnetic-field phasors, respectively, and the asterisk indicates complex conjugation. The corresponding instantaneous energy balance is

$$\nabla \cdot \boldsymbol{S}(t) + \frac{\partial u_{EM}(t)}{\partial t} = -\mathbf{J}(t) \cdot \boldsymbol{E}(t) \tag{2}$$

as obtained from Maxwell's equations [27]. In Eq. (2), $\boldsymbol{S}(t) = \boldsymbol{E}(t) \times \boldsymbol{H}(t)$ is the instantaneous Poynting vector, $u_{\mathrm{EM}}$ denotes electromagnetic energy density, and $\boldsymbol{J} \cdot \boldsymbol{E}$ represents the rate of work performed by the field on the specified current density $\mathbf{J}(t)$. For linear, nondispersive, lossless media with time-independent constitutive parameters, the conventional energy density is

$$\mathrm{u_{EM}} = (\mathbf{E} \cdot \mathbf{D} + \mathbf{B} \cdot \mathbf{H})/2 \tag{3}$$

with **D** and **B** denoting electric flux density and magnetic flux density, respectively [28, 29]. In dispersive or lossy media, stored energy and irreversible dissipation must be identified consistently with the constitutive model [30]. Unlike Eq. (1), all quantities in Eq. (2) are instantaneous; the time-averaged phasor expression cannot be substituted directly into the instantaneous balance. The two relations distinguish electromagnetic energy storage, energy transport across a boundary, and energy transferred from the fields to matter. The distinction is particularly important in reactive near-field WET, where energy may be stored and returned during successive excitation cycles without producing an equivalent net transfer to the receiver. Under the adopted phasor convention, the real part of the complex Poynting vector describes active power flux, whereas its imaginary part is associated with reactive exchange [28, 29]. The latter cannot generally be interpreted as an independent local stored-energy density. Large field amplitudes or substantial resonator energy therefore do not, by themselves, establish useful power delivery. Karalis et al. [27] established the theoretical basis for efficient non-radiative mid-range energy transfer through resonant coupling. Kurs et al. [31] subsequently demonstrated wireless power transfer using strongly coupled magnetic resonances, while Sample et al. [32] examined the corresponding circuit behavior, experimental performance, and range adaptation. These studies show that useful power delivery depends on coupling, resonator losses, and operating conditions rather than resonance alone. Resonance can facilitate energy exchange but cannot eliminate dissipation or guarantee efficient transfer under arbitrarily weak coupling. The associated coupling and loading mechanisms are examined in Section 2.2. The appropriate transfer model depends on the electrical dimensions of the source and receiver, their separation, and the operating wavelength. Near-field WET primarily exploits electric or magnetic coupling [27, 32], whereas radiative transfer carries energy through propagating electromagnetic waves, as reviewed for microwave and millimeter-wave power beaming by Rodenbeck et al. [6]. For a localized oscillating source, reactive field contributions generally decay more rapidly with distance than the leading radiative field, whose amplitude scales as $1/r$. Nevertheless, the reactive near field, radiating near field, and far field must be distinguished according to source electrical size and observation geometry rather than separation alone.

For an electrically large aperture of maximum dimension D, the conventional Fraunhofer criterion is [29, 33]

$$R \gtrsim \frac{2D^2}{\lambda} \tag{4}$$

where $R$ denotes the observation distance and $\lambda$ is the operating wavelength [29]. This criterion provides an approximate far-field condition rather than a universal boundary for every antenna configuration. Electrically compact links therefore require suitable field or circuit descriptions [27, 32], while finite-aperture propagation in the Fresnel region must account for the spatial field distribution instead of relying on an unqualified inverse-square approximation [6, 34].

Regardless of the propagation regime, received electromagnetic power must be defined with respect to a specified boundary. The time-averaged net power entering a receiver through a surface $A_{\mathrm{R}}$ is [28, 29]

$$P_{\mathrm{net,in}} = \iint_A \langle \mathbf{S}_{\mathrm{tot}} \rangle \cdot \hat{\mathbf{n}}_{\mathrm{in}} \, \mathrm{d}A \tag{5}$$

This expression follows from the surface-integral form of Poynting's theorem [28, 29], with the surface normal defined here as inward-directed. The total electromagnetic field is used to evaluate $\mathbf{S}_{\mathrm{tot}}$. For a closed surface, Eq. (5) gives the net electromagnetic power entering the enclosed volume. When applied to an open surface, it represents only the signed net flux through that surface and must not automatically be identified with incident or receiver-accepted power. Incident power requires a specified incoming field and receiving aperture. For antennas, the effective receiving aperture may differ from the physical aperture, while polarization, scattering, and impedance matching affect the relationship between incident radiation and power accepted at the receiving port [29]. Subsequent conversion introduces additional losses. Gu et al. [15] examined how nonlinear rectifier behavior and operating conditions influence RF-to-DC conversion in far-field wireless power harvesting. The useful electrical output therefore depends on both electromagnetic reception and the conversion process, rather than incident field intensity alone.

A representative direct radiative link can be described by the conceptual energy pathway

$$\mathrm{P_{source}} \rightarrow \mathrm{P_{EM}} \rightarrow \mathrm{P_{intercepted}} \rightarrow \mathrm{P_{accepted}} \rightarrow \mathrm{P_{electrical}} \rightarrow \mathrm{P_{load}}$$

This sequence illustrates the distinction between launched, intercepted, accepted, and useful power, which is essential when evaluating microwave power-beaming systems [6]. It is not, however, a universal set of efficiency reference planes. Reactive links need not contain a distinct interception stage [27], while relay-assisted and regenerative architectures introduce additional electromagnetic or electrical pathways, as discussed in Section 2.5.

For a specified load, the useful energy delivered during an operating interval is

$$E_{\mathrm{load}} = \int_{t_0}^{t_1} P_{\mathrm{load}}(t)\,\mathrm{d}t. \tag{6}$$

where $P_{\mathrm{load}}(t)$ denotes the instantaneous useful electrical power supplied to the load. Equation (6) defines the delivered energy over the selected observation interval. Meaningful comparisons consequently require consistent input and output boundaries, operating conditions, and time intervals. As illustrated by the distinction between transmitter output, intercepted RF power, and rectified electrical power in microwave power-beaming systems [6], an isolated component efficiency cannot represent the performance of an entire WET architecture.

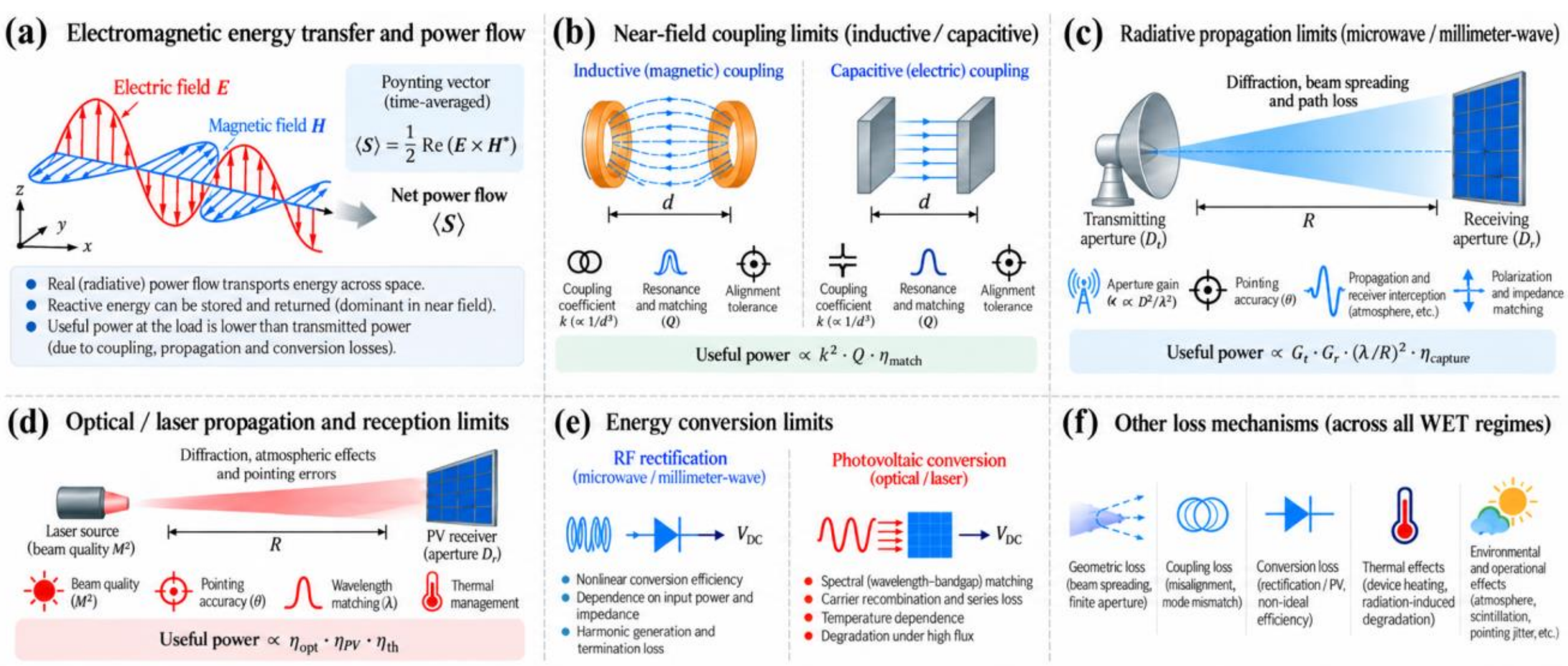

**Figure 2. Physical pathways and limiting processes in space WET.** (a) Electromagnetic energy transport and the distinction between net electromagnetic power flow and reactive energy storage and exchange, illustrated using the time-averaged Poynting vector. (b) Near-field inductive and capacitive coupling, highlighting the effects of coupling geometry, resonator quality factor, alignment, and impedance matching. (c) Microwave and millimeter-wave propagation, where wavelength, aperture dimensions,

separation, pointing, propagation conditions, and receiver interception jointly determine received power. (d) Optical propagation and reception, emphasizing diffraction, beam quality, pointing stability, wavelength matching, and receiver thermal constraints. (e) RF rectification and photovoltaic conversion, whose performance depends on incident power, impedance and electrical operating conditions, spectral matching, device losses, and temperature. (f) Cross-cutting geometric, coupling, conversion, thermal, environmental, and operational limitations. The dominant limitation depends on the transfer regime and operating conditions; useful load power reflects losses across the complete energy pathway.

The mechanisms summarized in Figure 2 differ primarily in how electromagnetic energy reaches the receiver and which processes limit its conversion into useful power. Near-field transfer is governed primarily by coupling strength, resonator quality factor, impedance matching, alignment, and loading, whereas radiative links introduce aperture-dependent beam formation, propagation, finite-aperture interception, pointing, and receiver-conversion constraints. For optical links, wavelength matching and receiver thermal conditions further influence photovoltaic conversion, while environmental and operational effects can modify several stages of the pathway. The figure should therefore be read as a causal map rather than a catalogue of technologies: large field amplitude, strong resonance, or high local conversion efficiency does not by itself establish high useful-load power because the controlling bottleneck can migrate among coupling, propagation, interception, conversion, thermal management, and downstream conditioning. Accordingly, useful-load power is determined by the complete energy pathway rather than by the performance of any single stage. These distinctions provide the basis for the mechanism-specific analyses in Sections 2.2–2.5.

The development of space WET has progressively addressed different parts of this energy-transfer pathway. Glaser proposed a foundational space solar-power architecture [11], while the Goldstone experiment demonstrated high-power microwave reception and rectification [12]. Subsequent mission studies, ground demonstrations, and flight experiments have investigated complementary capabilities in energy delivery, coupling, beam control, source conversion, and spacecraft integration [13, 34–40]. Figure 3 summarizes representative developments by chronology and evidence class, distinguishing concepts and mission studies from ground demonstrations, flight or orbital component experiments, and in-space transfer or system/interface demonstrations. The distributed space-energy architecture shown at the end of the timeline is prospective rather than an operationally demonstrated capability. Chronological progression does not imply a monotonic increase in power, technology readiness, or completeness of end-to-end validation. Their differing evidence scopes reinforce the need for consistent system-level assessment. Section 3 therefore maps the physical mechanisms and architecture-specific energy pathways onto the R0–R7 reference planes.

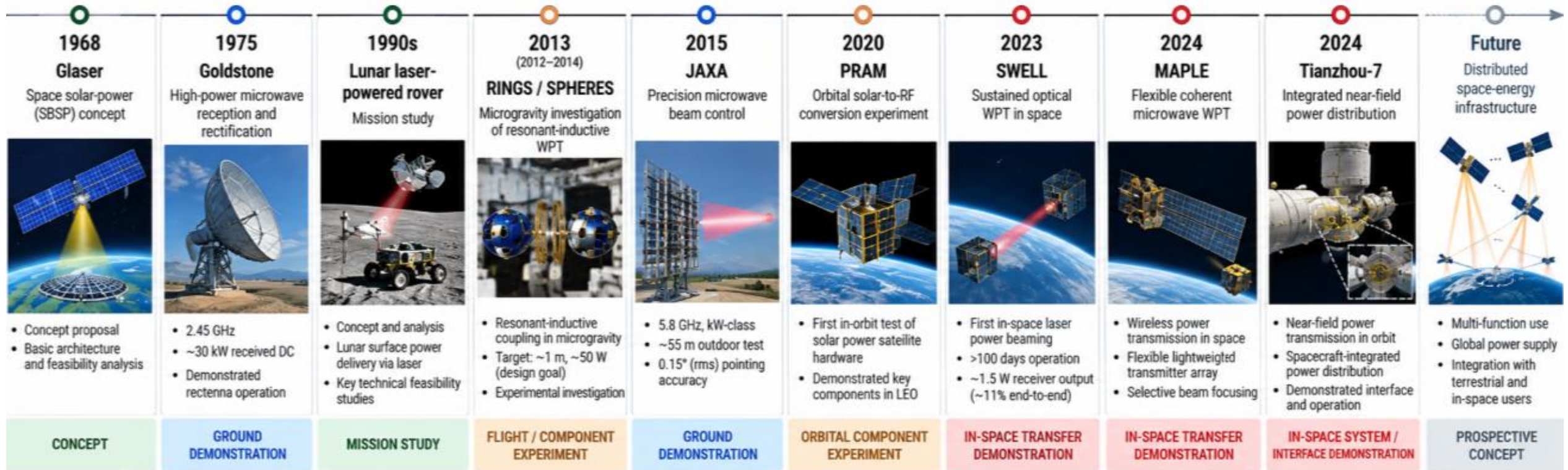

**Figure 3. Representative milestones and evidence levels in space WET.** Selected developments include Glaser's space solar-power concept [11], the Goldstone high-power microwave reception and rectification demonstration [12], a lunar laser-powered rover mission study [35], the RINGS/SPHERES microgravity investigation of resonant-inductive WPT [36], JAXA's ground-based precision microwave beam-control demonstration [37], PRAM's orbital solar-to-RF conversion experiment [38], SWELL's in-space optical power-transfer demonstration [39], MAPLE's flexible coherent microwave WPT experiment [13], and Tianzhou-7's spacecraft-integrated near-field power-distribution demonstration [40,41]. Evidence categories distinguish concepts and mission studies from ground demonstrations, flight or orbital component experiments, and in-space transfer or system/interface demonstrations. The distributed space-energy infrastructure shown at the far right is prospective and is included only to indicate a possible direction of architectural evolution. Chronological ordering does not imply a monotonic increase in transmitted power, technology readiness, or end-to-end energy-chain validation.

## 2.2 Near-Field WET: Coupling, Resonance, and Impedance Matching

Near-field WET exploits predominantly reactive electromagnetic coupling between closely spaced structures. In inductive power transfer (IPT), an alternating current in the transmitter produces a time-varying magnetic flux that induces an electromotive force in the receiver through Faraday induction. For a linear two-coil system, the mutual inductance, magnetic coupling coefficient (k), and induced receiver voltage are related by [27, 32]

$$M = k\sqrt{L_T L_R} \qquad V_R = -j\omega M I_T \tag{7}$$

where $L_T$ and $L_R$ are the transmitter and receiver self-inductances, respectively. And $\omega$ is the excitation angular frequency, $V_R$ is the induced receiver-voltage phasor and $I_T$ denotes the peak transmitter-current phasor. The voltage polarity follows the adopted winding and reference conventions. Greater separation or misalignment generally weakens magnetic flux linkage [42], although the induced voltage alone does not determine the power delivered to the load. The receiver also modifies the electrical conditions seen by the transmitter through reflected impedance. In a linear two-coil equivalent circuit [32, 43],

$$Z_{ref} = \frac{(\omega M)^2}{Z_R} \qquad P_L = \frac{1}{2}\left|I_{R,rms}\right|^2 R_L \tag{8}$$

where $Z_{ref}$ denotes the reflected impedance and $Z_R$ includes the receiver coil impedance, compensation network, and load; $I_{R,rms}$ is the peak receiver-current phasor, and $R_L$ is the resistive load. The reflected-impedance expression follows a consistent mutual-inductance circuit and port-current convention. Consequently, the operating point is governed by the interaction among magnetic coupling, source impedance, compensation, and loading rather than by mutual inductance alone [27, 32]. Resonant compensation reduces the net reactive impedance and facilitates energy exchange between coupled resonators, while resistive and other dissipative losses remain. For an ideal series RLC resonator, the resonance frequency and unloaded quality factor are

$$\omega_0 = \frac{1}{\sqrt{LC}} \qquad\qquad Q = \frac{\omega_0 L}{R} \tag{9}$$

where $\omega_0$ is the angular resonance frequency, $Q$ is the unloaded quality factor, and $L$, $C$, $R$ represent the resonator inductance, capacitance, and intrinsic loss resistance, respectively. The performance of two resonators tuned to a common resonance is determined by the competition between coupling and intrinsic dissipation [27], conveniently expressed through the dimensionless parameter $F = k^2 Q_T Q_R$, where $Q_T$ and $Q_R$ are their intrinsic unloaded quality factors. For an ideal linear two-resonator link operated at the common resonance, optimization of the receiver load yields [27, 32]

$$\eta_{\text{link,max}} = \frac{F^2}{\left(1+\sqrt{1+F^2}\right)^2} \tag{10}$$

Here, link efficiency is defined as the power dissipated in the receiver load divided by the power entering the modeled transmitter resonator. Equation (10) excludes source-conversion, power-electronic, and auxiliary losses and does not generally apply to arbitrary compensation topologies, detuned operation, or nonlinear rectification. The load that maximizes transfer efficiency does not necessarily maximize delivered power [27, 44]. Adjusting the receiver load changes both its current and the impedance reflected to the transmitter, thereby altering useful load dissipation and internal losses differently. The resulting optima depend on the source impedance, compensation topology, operating frequency, and imposed excitation conditions [32, 44]. At stronger coupling, frequency-dependent impedance and mode splitting may further separate the power and efficiency maxima [27, 44]. Although higher intrinsic quality factors reduce resonator dissipation, they can also increase sensitivity to frequency detuning. The usable bandwidth is therefore determined jointly by coupling, damping, loading, and matching [27, 32]. Moreover, observable spectral splitting depends on modal damping and the excitation and detection configurations; the appearance of two spectral peaks alone does not establish the operating frequency of maximum load power or efficiency. Experimental investigations have progressively addressed these limitations. Kurs et al. demonstrated approximately 60 W of nonradiative power transfer over distances exceeding 2 m, with a reported efficiency of approximately 40% [31]. Sample et al. subsequently investigated adaptive frequency tuning, reporting efficiencies above 70% over a 0–70 cm operating range [32], while Duong and Lee examined adaptive matching under misalignment [43]. These studies establish the relevance of maintaining favorable resonance and impedance conditions as coupling geometry changes. Their reported efficiencies, however, should be compared only after accounting for the respective measurement reference planes and operating conditions.

Capacitive power transfer (CPT) provides an alternative near-field mechanism in which energy is coupled through electric fields and displacement currents between electrodes [45, 46]. An alternating voltage establishes the coupling field, while the compensation network and downstream power electronics determine the electrical conditions for useful power delivery. For a lumped capacitance under fixed geometry, the electrode charge and current satisfy [45, 46]:

$$i_C(t) = C_c \frac{dv(t)}{dt} \qquad |X_C| = \frac{1}{\omega C_c}. \tag{11}$$

where $i_C$ is the coupling-capacitor current, $C_c$ is the defined coupling capacitance, $X_C$ is the magnitude of its capacitive reactance, and $v$ is the voltage across the defined capacitance. For an ideal parallel-plate coupler [45, 46]:

$$C_c \simeq \frac{\varepsilon A_{eff}}{d} \tag{12}$$

with $\varepsilon$, $A_{\text{eff}}$, and $d$ denoting dielectric permittivity, effective electrode overlap area, and separation, respectively. This approximation neglects fringing fields and interactions with neighboring conductors. When the coupling geometry varies with time, the current becomes

$$i_c(t) = \frac{d[C_c(t)v(t)]}{dt} = C_c(t)\frac{dv(t)}{dt} + v(t)\frac{dC_c(t)}{dt} \tag{13}$$

provided that a lumped, quasistatic capacitance description remains applicable [45, 46]. Unlike an isolated two-terminal capacitor, a practical CPT interface must provide a complete displacement-current return path. A differential four-plate coupler, for example, contains six pairwise capacitances associated with the principal transfer paths, cross-coupling, and same-side electrode interactions [46, 47]. These contributions are captured by the Maxwell capacitance matrix [46, 47],

$$\mathbf{q} = \mathbf{C}_{\boldsymbol{M}}\boldsymbol{v} \tag{14}$$

where $\boldsymbol{q}$ and $\boldsymbol{v}$ are the electrode charge and potential vectors, respectively, defined under a specified potential reference and electrostatic boundary condition, and $\mathbf{i}$ is the corresponding electrode-current vector. For fixed geometry and time-invariant dielectric properties,

$$\mathbf{i} = \mathbf{C}_{\boldsymbol{M}}\frac{d\mathbf{V}}{dt} \tag{15}$$

Misalignment, dielectric variation, and nearby conductive structures modify the capacitance network and displacement-current distribution, with corresponding changes in effective port admittance, common-mode currents, stray fields, and delivered power [45–47]. Reduced two-capacitor models are appropriate only when the neglected couplings have a sufficiently small influence under the specified geometry and boundary conditions [46, 47]. Although compensation can mitigate capacitive reactance [45, 46], it cannot recover geometric coupling lost through misalignment or eliminate dielectric dissipation, voltage stress, and stray-field constraints. A related lunar-surface concept investigated the use of lunar sand to expand the area over which wireless power can be supplied to mobility vehicles [48]. This concept should be treated as a distinct surface-mediated electric-field power-transfer approach rather than as evidence that conventional plate-based CPT is insensitive to geometry or dielectric losses.

For spacecraft applications, relative motion and misalignment can perturb near-field coupling [36, 42], while variable loading changes the impedance condition seen by the transmitter [32, 43]. Thermal deformation may introduce additional geometry changes that require mission-specific assessment. Beyond these operating-point variations, the reliability of power electronics and their performance under mission-relevant environmental conditions require separate assessment. Tang et al. demonstrated that γ-ray irradiation can modify the electrical characteristics ofGaN power devices, while the efficiency of their tested WPT system remained largely unchanged under the investigated conditions [49]. Radiation-hardness assurance for spaceborne electronic components must account for total ionizing dose, displacement damage, and single-event effects in relation to the specified mission environment [50, 51]. Spacecraft charging, dielectric breakdown, and charge accumulation on electrically floating conductors introduce additional risks that require dedicated assessment, particularly for exposed CPT electrodes and insulating structures [52]. Thermal-vacuum verification and functional testing under specified environmental and operating conditions are also necessary within the applicable spacecraft equipment test programme [53]. Accordingly, near-field WET should be evaluated in terms of delivered load power, efficiency, and operating stability across representative alignment, thermal, electrical, and environmental conditions. Terrestrial link efficiency alone does not establish qualification for space operation.

### 2.3 Microwave and Millimeter-Wave WET: Beam Formation, Propagation, Capture, and Rectification

Microwave and millimeter-wave WET delivers power through propagating electromagnetic fields. Unlike near-field coupling, its performance depends on the spatial distribution of the transmitted field, its interception by the receiver, RF acceptance, and subsequent conversion into DC electricity. For a transmitting aperture of physical area $A_\mathrm{t}$, the far-field gain is related to its effective aperture by [54, 55]

$$G_t \simeq \frac{4\pi\eta_{a,t}A_{e,t}}{\lambda^2} = \frac{4\pi\eta_{a,t}A_t}{\lambda^2} \quad (16)$$

Where $G_t$ denotes the transmitting-antenna gain, $\lambda$ is the wavelength, $A_\mathrm{e,t}$ is the effective aperture, $A_t$ is the physical transmitting-aperture area.and $\eta_\mathrm{a,t}$ is the transmitting-aperture efficiency defined consistently with the antenna-gain convention. Directional radiation can also be synthesized through coherent excitation of an antenna array. In an idealized scalar representation, the field at an observation point $\boldsymbol{r}$ is expressed as [34, 55]

$$E(\mathbf{r}) \propto \sum_{n=1}^{N} \frac{a_n}{r_n} \exp[j(\phi_n - kr_n)] \quad (17)$$

where E($\mathbf{r}$) denotes the complex scalar field at the observation point, $a_n$and $\phi_n$ are the equivalent field amplitude and imposed phase of element $n$, respectively, $r_n$is its distance to the observation point, and $k = 2\pi/\lambda$ is the free-space wavenumber; n = 1,…,N indexes the N array elements, and j = $\sqrt{-1}$. Appropriate phase compensation produces constructive interference at the intended location. A linear phase gradient steers a far-field beam, whereas a focusing phase profile accounts for the propagation distance to a finite-range target [34, 55]. Such control redistributes radiated power spatially without increasing the total power generated by the array. Equation (17) illustrates the interference mechanism rather than a complete array model; quantitative designs must additionally consider element radiation patterns, polarization, mutual coupling, feed-network losses, and phase errors [6, 55]. The synthesized field subsequently evolves through diffraction and propagation. Increasing the electrical aperture $D/\lambda$ reduces the characteristic beam divergence, which scales approximately as $\lambda/D$, but also tightens pointing requirements. At a separation $R$, a small angular pointing error $\delta\theta$ produces an approximate transverse displacement $R\delta\theta$. For antennas operating in one another's far field under aligned, polarization-matched, unobstructed free-space conditions, the available received RF power follows the Friis transmission relation [54]:

$$P_\mathrm{r} = P_\mathrm{t} G_\mathrm{t} G_\mathrm{r} \left(\frac{\lambda}{4\pi R}\right)^2 \tag{18}$$

Here, $P_\mathrm{t}$ is the RF power accepted by the transmitting antenna, $P_\mathrm{r}$ is the available RF power at the receiving antenna terminals under a conjugate-matched reference, and $G_\mathrm{t}$ and $G_\mathrm{r}$ are consistently defined antenna gains. Pointing, polarization, impedance mismatch, and feed losses must be assigned to the appropriate power boundaries without double counting. For electrically large apertures operating in the Fresnel region, the far-field approximation is insufficient. Pinero et al. [34] investigated microwave power transmission between circular apertures at arbitrary distances, demonstrating the importance of finite-aperture propagation, illumination taper, and wavefront curvature in determining transfer efficiency. Such links therefore require field-based propagation analysis rather than direct extrapolation of Eq. (18).

The propagated field determines the fraction of radiated power intercepted by the receiving aperture. For a specified receiving surface $A_\mathrm{R}$, the interception fraction is

$$\eta_\mathrm{int}(\mathbf{\Delta}) = \frac{P_\mathrm{int}(\mathbf{\Delta})}{P_\mathrm{rad}} = \frac{1}{P_\mathrm{rad}} \iint_{A_\mathrm{R}(\mathbf{\Delta})} \langle \mathbf{S}_\mathrm{inc} \rangle \cdot \hat{\mathbf{n}} \, \mathrm{d}A \tag{19}$$

where $P_\mathrm{rad}$ is the total radiated power, $P_\mathrm{int}$is the incident power intercepted by the receiver, $\langle \mathbf{S}_\mathrm{inc} \rangle$ is the time-averaged Poynting vector of the specified incident field, $\hat{\mathbf{n}}$ is the receiving-surface normal oriented toward the incident power flow, and $\mathbf{\Delta}$ denotes lateral displacement between the beam and receiving aperture. This field-based interception formulation is consistent with the finite-aperture treatment investigated by Pinero et al. [34]. It assumes a prescribed incident field; significant scattering or near-field interactions require a more complete vector-field treatment. As established in Section 2.1, intercepted power is distinct from RF power available at the receiving antenna terminals. Receiver polarization, radiation characteristics, and impedance matching determine the subsequent transfer to the electrical interface [54, 55]. High on-axis intensity or nominal beam capture therefore does not necessarily translate into high delivered RF power.

These relations reveal the coupled dependence on aperture, range, and frequency. The receiving effective aperture is related to antenna gain by [54]

$$A_t = \frac{G_r \lambda^2}{4\pi} \tag{20}$$

Substituting the effective-aperture relations into Eq. (18) gives the ideal far-field result [54]:

$$\frac{P_r}{P_t} = \frac{A_{e,t} A_{e,r}}{\lambda^2 R^2} = \frac{\eta_{e,t} \eta_{a,r} A_t A_r}{\lambda^2 R^2} \tag{21}$$

For fixed physical transmitting and receiving apertures with approximately constant aperture efficiencies, the ideal geometric coupling therefore scales as $1/(\lambda^2 R^2)$. A different frequency dependence arises when antenna gains, rather than physical areas, are held fixed. Neither scaling can be extrapolated beyond the physical capture limit or applied indiscriminately to Fresnel-region links [34]. Aperture dimensions and operating frequency must consequently be selected jointly with propagation distance, pointing uncertainty, aperture mass, and RF-generation and distribution losses [55].

After RF reception, a rectenna produces DC power through an impedance-matching network and a nonlinear rectifying circuit [56, 57]. The nonlinear current–voltage response of the rectifying device generates a DC component, while the matching network, DC load, temperature, and harmonic terminations influence the conversion process [56, 57]. Defining $P_{\mathrm{RF,in}}$ as the RF power accepted at the rectifier input, after any preceding matching-network losses, the RF-to-DC conversion efficiency is

$$\eta_{\mathrm{RF-DC}} = \frac{P_{\mathrm{DC}}}{P_{\mathrm{RF,in}}} = f\left(P_{\mathrm{RF,in}}, R_L, T, \mathcal{H}\right) \tag{22}$$

where $P_{\mathrm{DC}}$ is the DC power delivered to the defined DC load, $R_{\mathrm{L}}$is the DC load, $T$ is temperature, and $\mathcal{H}$ represents the harmonic termination conditions. McSpadden et al. [56] investigated the theoretical and experimental performance of a rectenna element, providing a basis for treating rectification as a device- and operating-point-dependent process. Because rectifier input impedance and conversion efficiency vary with operating conditions [15, 16, 56], nonuniform illumination and pointing-induced power fluctuations can alter both matching and DC conversion.

Under a quasi-static approximation, the relevant time-averaged DC output is

$$\overline{P}_{\mathrm{DC}} = \frac{1}{t_1 - t_0}\int_{t_0}^{t_1} \eta_{\mathrm{RF-DC}}\left[P_{\mathrm{RF,in}}(t), R_L, T, \mathcal{H}\right] P_{\mathrm{RF,in}}(t)\, dt \tag{23}$$

where $t_0$ and $t_1$ delimit the averaging interval. Owing to the nonlinear dependence of conversion efficiency on input power [56, 57], this result generally differs from the value obtained by evaluating rectifier efficiency at the mean RF input power. If circuit or control dynamics are significant, time-domain analysis is required instead of a static efficiency map. Beam capture, peak rectifier efficiency, and average DC output are therefore distinct optimization objectives. Matching and loading should be designed for the expected distribution of received power rather than a single nominal operating point.

Experimental demonstrations illustrate both the feasibility of radiative WET and the importance of distinguishing measurement boundaries. The 1975 Goldstone experiment transferred microwave power over 1.54 km at 2.388 GHz and delivered up to 30.4 kW of DC power [12], with a reported receiver-array collection–conversion efficiency exceeding 80% [12]. At higher frequencies, McSpadden et al. [56] investigated a 35 GHz rectenna and reported an RF-to-DC conversion efficiency of 29% at an input power of 120 mW. More recently, Sato et al. [58] demonstrated 2.9 W of DC output from 14 W of transmitter RF output in a K-band phased-array link, corresponding to a transmitter-RF-to-receiver-DC power ratio of 20.7%. The reported local rectification efficiency of 73.9% [58] characterizes a different power boundary. These measurements describe receiver-array collection and conversion, individual rectifier performance, and transmitter-RF-to-receiver-DC transfer, respectively, and should not be compared as equivalent end-to-end efficiencies.

For spacecraft applications, the design objective is useful DC delivery under mission-specific constraints rather than maximum antenna gain or conversion efficiency in isolation. Aperture dimensions and mass, propagation geometry, pointing statistics, RF-generation and feed losses, thermal dissipation, and receiver operating conditions jointly determine system performance [6, 55]. Koert and Cha [59] examined millimeter-wave technology for space power beaming, providing a basis for considering higher-frequency operation in space-energy architectures. Millimeter-wave frequencies can reduce the physical aperture required for a specified beamwidth, but this geometric advantage improves delivered DC power only when the capture benefit exceeds the associated generation, phase-control, distribution, and rectification penalties [55, 59]. The complete beam-to-DC pathway should therefore be evaluated over representative operating conditions rather than optimized through isolated antenna or rectifier metrics.

### 2.4 Optical WET: Diffraction, Beam Quality, and Photon-to-Electric Conversion

Optical WET follows the finite-aperture propagation and capture principles established for microwave power beaming, but its shorter wavelength shifts the principal design constraints toward beam quality, pointing stability, wavelength-selective absorption, and photovoltaic conversion. For an ideal paraxial Gaussian beam, the $1/e^2$ irradiance radius at a propagation distance $z$ from the beam waist is given by [60]

$$\omega(z) = \omega_0\sqrt{1 + (\frac{z}{z_R})^2} \tag{24}$$

where $w_0$is the beam-waist radius and $z_R$is the Rayleigh range, defined as [60]

$$z_R = \frac{\pi w_0^2}{\lambda} \tag{25}$$

The corresponding far-field half-angle divergence is [60]

$$\theta \simeq \frac{\lambda}{\pi w_0} \tag{26}$$

where $\lambda$ denotes the optical wavelength. Consequently, Eqs. (24)– (26) describe the diffraction-limited relationship among beam waist, propagation distance, and beam spreading. Real laser beams depart from this ideal behavior because of higher-order spatial modes and wavefront imperfections. Using consistently defined beam-waist width and far-field divergence, the beam-quality factor $M^2$ gives [61]

$$\theta \simeq \frac{M^2\lambda}{\pi\omega_0} \tag{27}$$

where $M^2 = 1$ represents an ideal fundamental Gaussian beam and $M^2 > 1$ indicates a larger beam-parameter product relative to the diffraction-limited reference. Johnston [61] developed a measurement approach for determining $M^2$ from beam propagation characteristics. Although Eq. (27) quantifies the relationship between beam width and divergence, it does not specify the complete irradiance distribution of a non-Gaussian beam. Wavefront aberrations, aperture truncation, and beam shaping therefore require a suitable propagation model or measured spatial irradiance profile. For non-axisymmetric beams, the propagation parameters should be evaluated separately along the relevant transverse axes.

The propagated irradiance distribution determines the optical power intercepted by the receiving aperture. For a centered ideal Gaussian beam incident normally on a coaxial circular aperture of radius $a$, integration of the Gaussian irradiance profile described by [60] yields

$$\eta_{int} = \frac{P_{int}}{P_{beam}} = 1 - exp\left(-\frac{2a^2}{w^2(z)}\right) \tag{28}$$

where $P_{\text{beam}}$ is the total optical power crossing the receiving plane and $P_{\text{int}}$ is the power incident within the specified aperture. Equation (28) describes geometric interception and excludes subsequent reflection, absorption, and photovoltaic conversion losses. For lateral displacement, aperture truncation, or non-Gaussian illumination, interception must instead be obtained by integrating the actual irradiance distribution over the receiver surface. A smaller beam footprint can increase nominal capture under perfect alignment, but it also reduces tolerance to pointing errors. Receiver sizing and beam control must consequently account for the expected beam-position distribution rather than relying exclusively on diffraction-limited spot size.

Once the optical power reaches the receiver, electrical conversion proceeds through photon absorption, photocarrier generation, carrier collection, and power extraction. The energy of an incident photon is [62]

$$E_{ph} = \frac{hc}{\lambda} \tag{29}$$

where $h$ is Planck's constant, $\nu$ is optical frequency, and $c$ is the speed of light. In a conventional photovoltaic absorber, photons with energies below the semiconductor bandgap cannot generate band-to-band electron–hole pairs through ordinary single-photon absorption. Photons with sufficient energy may generate carriers, although not all absorbed optical energy becomes extractable electrical energy. Algora et al. [62] reviewed photovoltaic laser power converters and identified spectral matching, absorption, carrier recombination, and device-level electrical losses as important determinants of conversion performance. Below-bandgap transmission, parasitic absorption, incomplete absorption above the bandgap, and carrier thermalization limit optical-energy utilization. Wavelength–bandgap matching can reduce spectral mismatch and thermalization losses, although operation near the absorption edge may require a greater optical path length to maintain adequate absorption [62]. Photogenerated carriers must subsequently be collected before recombination, while series resistance and shunt leakage modify the illuminated current–voltage characteristics [62]. The extractable electrical power therefore depends on the absorber structure and operating conditions as well as the incident wavelength and optical power.

For a photovoltaic receiver operating at its maximum power point, the electrical output is

$$P_{\text{MPP}} = V_{\text{MPP}} I_{\text{MPP}} \tag{30}$$

where $V_{\text{MPP}}$ and $I_{\text{MPP}}$ denote the voltage and current at the maximum power point. The importance of device architecture, operating temperature, and power extraction in photovoltaic laser converters is discussed by Algora et al. [62]. In an array receiver, spatially nonuniform illumination can create photocurrent mismatch among series-connected cells, reducing extractable power even when total intercepted optical power remains unchanged. The electrical response must therefore be evaluated using the actual illumination distribution and cell-interconnection configuration. At high optical power densities, this requirement becomes closely coupled to thermal management.

Photovoltaic conversion and heat rejection cannot be treated independently. Optical energy absorbed by the receiver but not exported as electrical power contributes to its thermal load, subject to the specified energy balance. Temperature-dependent changes in photovoltaic characteristics can reduce voltage and fill factor, while the magnitude of these effects depends on the device material and operating conditions [62]. In a vacuum environment, external convective cooling is absent; sustained operation consequently depends on conductive heat transport and radiative heat rejection. For a steady-state photovoltaic receiver with no additional significant exported energy channels, conservation of energy gives

$$P_{abs} - P_{elec} = Q_{cond} + \mathrm{Q} \tag{31}$$

where $P_{\mathrm{abs}}$ is the optical power absorbed within the specified receiver boundary, $P_{\mathrm{elec}}$ is the exported electrical power, and $Q_{\mathrm{cond}}$ and $Q_{\mathrm{rad}}$ are the net conductive and radiative heat-rejection rates, respectively. Transient operation requires an additional thermal-energy storage term. Because temperature affects photovoltaic performance, increasing optical concentration does not necessarily maximize sustained electrical output. Beam distribution, cell interconnection, electrical operating point, and heat rejection must instead be evaluated through a coupled optical–electrical–thermal model.

Ochiai et al. [63] demonstrated a 1-km atmospheric optical WET link using a 1070 nm laser with 1035 W of transmitted optical power. Their $600 \times 600\ \mathrm{mm}^2$ crystalline-silicon photovoltaic receiver produced approximately 152 W of electrical output [63]. Transmitter-side beam shaping and receiver-side homogenization improved illumination uniformity, while capacitive smoothing reduced output fluctuations [63]. The ratio of receiver electrical output to transmitted optical power was approximately 15%. This measurement includes propagation, interception, and photovoltaic conversion losses under the reported conditions but excludes the electricity consumed in generating the laser beam. It therefore represents a defined optical-to-electrical transfer efficiency rather than complete electrical-input-to-electrical-output efficiency.

For exoatmospheric WET, the absence of atmospheric absorption and turbulence removes two terrestrial propagation constraints, whereas diffraction, wavefront errors, platform jitter, and receiver heating remain. The Gaussian propagation relations [60] and beam-quality characterization [61] provide a basis for assessing the optical aperture and beam-divergence requirements. Smaller diffraction-limited apertures are possible at optical wavelengths for a specified beamwidth, but useful power delivery also depends on maintaining adequate interception and wavelength-matched photovoltaic conversion [62]. Thermal rejection and downstream power conditioning impose additional constraints on sustained operation. Optical WET architectures must therefore be evaluated from laser electrical input through propagation and photovoltaic conversion to regulated load output, using the consistent power boundaries established in Section 3.

### 2.5 Wavefront Control, Reconfigurable Apertures, Relays, and Cross-Regime Selection

Wavefront control extends WET beyond fixed source–receiver geometries by modifying the spatial distribution of electromagnetic power through engineered aperture responses. Metasurface theory provides a framework for describing how spatially varying electromagnetic boundary conditions can redirect or reshape an incident field [64]. For a locally responding transmissive aperture, the complex transmission coefficient can be represented by

$$t(\boldsymbol{r}_\perp) = a(\boldsymbol{r}_\perp)\exp[j\phi(\boldsymbol{r}_\perp)] \tag{32}$$

and the transmitted field is approximated by

$$E_{\text{out}}(\boldsymbol{r}_\perp) = t(\boldsymbol{r}_\perp)E_{\text{in}}(\boldsymbol{r}_\perp) \tag{33}$$

Here, $\boldsymbol{r}_\perp$ denotes the transverse position, while $a$ and $\phi$ represent the local field-amplitude transmission coefficient and phase response, respectively. Equations (32) and (33) provide a simplified description of the aperture response; a complete electromagnetic treatment must account for the relevant field components, boundary conditions, and polarization channels [64, 65].

With the spatial phase convention $E \propto \exp(j\boldsymbol{k} \cdot \boldsymbol{r})$, and with $\nabla_\perp$ denoting the gradient in the transverse plane the local phase-gradient approximation gives [64, 65]

$$\boldsymbol{k}_{\parallel,\text{out}} - \boldsymbol{k}_{\parallel,in} = \nabla_\perp\phi \tag{34}$$

where the subscripts denote transverse wavevector components. A transverse phase gradient redirects the outgoing field, whereas a focusing profile compensates propagation-phase differences across the aperture to concentrate power at a finite-distance receiver. Li et al. [8] demonstrated electronically controlled WPT routing using a reconfigurable transmissive metasurface. Unlike source-array phase control, an intermediate aperture operates on an incident field and remains constrained by the electromagnetic power available at its surface.

Wavefront manipulation redistributes electromagnetic power rather than generating additional energy. Asadchy et al. [65] showed that local phase control alone does not guarantee arbitrary ideal reflection or refraction; particular transformations require appropriate bianisotropic or nonlocal electromagnetic responses. Their analysis also demonstrates that electric-field amplitude may change across a lossless surface even when the transmitted power is conserved. Accordingly, the field-amplitude coefficient $a$ cannot generally be identified with the square root of power transmittance without suitable impedance and modal normalization. Polarization conversion, angular dependence, mutual coupling, and finite-aperture diffraction impose further restrictions on the local scalar approximation [64]. Wavefront optimization should therefore target power delivered to the finite receiver rather than local field enhancement alone. An intermediate aperture or resonant structure can exploit this redistribution to modify the energy-transfer path. A passive relay redirects intercepted electromagnetic power through scattering, transmission, or resonant coupling without supplying net electromagnetic gain. In two-dimensional magnetic-resonator arrays, Morita et al. [9] investigated how relay loading and routing geometry affect transfer efficiency. Their surface-routing method accounts for cross-coupling between resonators, with simulations indicating efficiency improvements of up to 10% relative to linear routing for the investigated 1.5m transfer configurations. This result illustrates the importance of route-dependent coupling but should not be interpreted as a universal efficiency gain for relay-assisted WET. An electronically reconfigurable aperture may require bias and control power while remaining electromagnetically passive with respect to the transferred signal. Amplifying relays differ by drawing on external electrical energy to increase their outgoing electromagnetic power. Regenerative relays introduce an electrical intermediate stage: received electromagnetic energy is converted, potentially stored, and subsequently used to generate a new carrier. Del Prete et al. [66] experimentally demonstrated this mechanism in a battery-less 2.45 GHz relay node that switches between rectification and oscillation. Their prototype stores harvested energy before using it to support RF generation, establishing a distinction between energy regeneration and passive electromagnetic redirection. For a passive relay operating at steady state without an additional energy source, conservation of energy gives

$$P_{in,EM} = P_{out,EM} + P_{loss} \tag{35}$$

where $P_{\mathrm{in,EM}}$ includes all incoming electromagnetic power crossing the specified relay boundary, $P_{\mathrm{out,EM}}$ includes all outgoing electromagnetic channels, and $P_{\mathrm{loss}}$ denotes dissipated power. Equation (35) assumes negligible change in stored energy over the averaging interval. The power redirected towards the intended receiver is only a subset of the total outgoing power; reflection and parasitic scattering must therefore be included when evaluating the relay balance.

For an amplifying relay, the corresponding steady-state balance becomes

$$P_{in,EM} + P_{ext} = P_{out,EM} + P_{loss} \tag{36}$$

where $P_{\mathrm{ext}}$ denotes external electrical power entering the specified relay boundary. If the system is transient, an additional energy-storage term is required. Auxiliary control loads outside this boundary must be included when the comparison is extended to the complete architecture.

The regenerative pathway is more appropriately expressed in terms of energy when reception and retransmission occur at different times:

$$E_{int} \xrightarrow{\eta_{rec}} E_{rec} \longrightarrow U_{store} \xrightarrow{\eta_{tx}} E_{out,EM} \tag{37}$$

Here, $E_{\mathrm{int}}$ is the electromagnetic energy intercepted during the receiving interval, $E_{\mathrm{rec}}$ is the recovered electrical energy, and $U_{\mathrm{store}}$ represents the intermediate stored energy. The arrows describe the conversion and storage sequence rather than a complete conservation equality. Del Prete et al. [66] measured the charging of a storage capacitor during rectification and its subsequent discharge during oscillator operation, demonstrating why storage dynamics cannot be omitted from the description of such a relay. External inputs, auxiliary consumption, storage losses, and changes in stored energy must be accounted for in the full energy balance.

Despite introducing additional losses, a passive relay can improve delivered power when it bypasses blockage or provides more favorable coupling than the direct path. The improvement arises from recovering otherwise poorly coupled power, not from increasing the electromagnetic energy available at the relay. For a simplified sequential two-hop pathway, let $P_{\mathrm{S}}$ denote source-launched electromagnetic power, $P_{\mathrm{R,int}}$ relay-intercepted power, $P_{\mathrm{R,out}}$ power launched into the designated relay-to-receiver channel, $P_{\mathrm{F,int}}$ power intercepted by the final receiver, and $P_{\mathrm{L}}$ useful load power. The pathway can then be written as

$$P_L = P_s\left(\frac{P_{R,int}}{P_s}\right)\left(\frac{P_{R,out}}{P_{R,int}}\right)\times\left(\frac{P_{F,int}}{P_{R,out}}\right)\left(\frac{P_L}{P_{F,int}}\right) \tag{38}$$

Each ratio corresponds to successive, nonoverlapping power boundaries. Equation (33) is a power-accounting identity for the defined sequential pathway, rather than an independently predictive efficiency model. Its use as a product of separately evaluated hop efficiencies requires the interactions between stages to be represented consistently. Morita et al. [9] showed that cross-coupling in two-dimensional resonator arrays complicates independent route analysis and that additional relay-loading conditions can be used to control the current distribution. Such systems require a coupled-circuit or electromagnetic model when the relay interactions cannot be neglected.

A related limitation arises when multiple coherent fields illuminate the same receiver. Their complex amplitudes must be combined before the accepted power is evaluated:

$$\boldsymbol{E}_{tot} = \boldsymbol{E}_{direct} + \boldsymbol{E}_{relay} \tag{39}$$

Krasnok et al. [67] demonstrated that an auxiliary coherent signal injected through a receiving antenna's load port can alter the local wave impedance and improve energy extraction under mismatched conditions. Their analysis explicitly subtracts the auxiliary input when evaluating net transferred power. Although this configuration differs from a passive two-hop relay, it establishes the importance of interference, matching, and consistent input–output accounting. Consequently, coherent direct and relayed paths cannot generally be assessed by adding independently calculated received powers without their interference terms.

The relevant comparison is the useful power delivered to the same final load under equivalent source-power, receiver, and operating conditions:

$$\Delta P_L = P_L^{relay} - P_L^{direct} \tag{40}$$

A positive $\Delta P_{\mathrm{L}}$ indicates an improvement in delivered load power under the specified conditions, but does not establish higher architecture-wide efficiency. Auxiliary consumption, additional conversion losses, and changes in stored energy may offset the gain. For time-varying links, the corresponding delivered energies must be compared over an identical operating interval.

The dominant physical limitation provides a more useful basis for cross-regime selection than operating frequency or transfer distance alone. Near-field links depend strongly on geometry-dependent coupling, resonator dissipation, and impedance matching [36, 42]. Microwave and millimeter-wave links are constrained by finite-aperture interception, pointing stability, and nonlinear RF-to-DC conversion [15, 54, 56]. Optical links couple diffraction and beam quality [60, 61] to wavelength-dependent photovoltaic conversion and receiver thermal conditions [62]. Wavefront control and relay structures offer additional flexibility when direct coupling is unfavorable, provided that improved geometric access compensates for interception, redirection, conversion, and control losses.

For spacecraft applications, relative motion and deployment tolerances can perturb near-field coupling, while attitude jitter and changing separation alter radiative beam interception. Aperture mass, available electrical power, and heat-rejection capacity further constrain feasible operating points. The terrestrial demonstrations discussed above establish relevant transfer mechanisms but do not, by themselves, demonstrate their qualification for the space environment. Architecture selection must therefore evaluate useful load power together with the resources and operating conditions required to sustain it. The R0–R7 reference-plane framework introduced in Section 3 provides a consistent basis for extending these mechanism-level comparisons to complete WET architectures.

## 3. Energy Pathway and System Architecture

Although Section 2 establishes the physical mechanisms governing WET, comparing the performance of different architectures requires a further distinction: reported efficiencies may refer to different portions of the energy-transfer pathway. Rodenbeck et al. [6] reviewed microwave and millimeter-wave power-beaming systems, for which launched electromagnetic power and receiver-side power collection constitute distinct performance boundaries. Gu et al. [15] examined nonlinear rectification and rectenna design, highlighting the operating conditions that govern RF-to-DC conversion. Consequently, transmitter conversion, beam collection, receiver electrical output, and useful regulated load power cannot be treated as interchangeable measures of system performance. A meaningful comparison must first establish which input and output boundaries are included in each reported result. To address this issue, this Review introduces an R0–R7 reference-plane framework that traces power from architecture-level electrical input to useful regulated load output. Rather than prescribing a universal efficiency metric, the framework provides a consistent basis for identifying the energy-transfer stages included in a measurement and those remaining outside its reported boundary. Figure 4 maps these reference planes onto direct, redirected, active, regenerative, and multi-node WET architectures, distinguishing their energy pathways and the additional inputs or intermediate conversion stages introduced by different system configurations. Section 3.1 formally defines the reference planes and associated efficiency expressions, while Sections 3.2 and 3.3 extend the accounting approach to intermediate energy-routing architectures and multi-node systems, respectively.

### 3.1 Energy Reference Planes and Direct Transfer

A direct WET architecture transfers energy from a source to a receiver without an independently located intermediate routing stage. Its reference-plane sequence is

$$R0 \rightarrow R1 \rightarrow R4 \rightarrow R5 \rightarrow R6 \rightarrow R7$$

Here, R0 denotes attributable architecture-level electrical input, including relevant auxiliary consumption; R1, electromagnetic power launched into the specified transfer mode; R4, power incident on the designated receiving aperture; R5, power accepted or coupled by the receiver; R6, converted electrical output; and R7, useful regulated load power. R2 and R3 are omitted because no separate routing stage is present. These functional definitions apply across WET modalities, although their physical implementation and measurement methods differ.

The efficiency between two declared reference planes is defined as

$$\eta(R_i \rightarrow R_j) = \frac{P(R_j)}{P(R_i)} \tag{41}$$

where $P(R_i) > 0$, and both powers are evaluated under consistent operating conditions and compatible measurement intervals. Microwave power-beaming studies distinguish launched and received electromagnetic power [6], whereas rectenna research addresses receiver-side conversion performance [15]. Equation (41) formalizes the reference-plane convention adopted here. For a direct link, the R0-to-R7 ratio represents end-to-end electrical efficiency only when all relevant inputs and changes in stored energy are properly accounted for; a narrower ratio characterizes the specified transfer segment.

The importance of these boundaries is evident in experimental results. Sato et al. [58] reported a K-band link producing 2.9 W of DC output from 14 W of transmitter RF power, giving

$$\eta(R1 \rightarrow R6) = 20.7\%$$

Its reported rectifier efficiency of 73.9% [58] refers to a narrower receiver-side boundary. Dickinson's Goldstone experiment [12] produced 30.4 kW of DC power and reported a peak collection–conversion efficiency exceeding 80%; because this metric uses a receiver-side denominator, it is not directly comparable with Sato's R1-to-R6 efficiency. The reported values describe different portions of the energy pathway rather than conflicting measures of the same performance.

For space WET, three distinctions are particularly relevant. R0 and R1 separate electrical input from launched electromagnetic power, accounting for source conversion, feed networks, power combining, and auxiliary consumption. Brown [68] examined the components of microwave power transmission for solar power satellites, while laser WPT efficiency was studied [7] address the corresponding optical source-conversion losses. R4 and R5 distinguish incident power from receiver-accepted power, which may be reduced by reflection, polarization or impedance mismatch, and combining losses [15]. Finally, R6 and R7 separate converted electrical output from useful load power after regulation and downstream processing. Storage losses and changes in stored energy must be included whenever storage lies within the declared system boundary.

The framework also clarifies the scope of different demonstrations. Ochiai et al. [63] reported approximately 150 W of receiver electrical output from 1035 W of transmitted optical power over a 1 km laser link, establishing an R1-to-R6 measurement boundary. Mardani et al. [69] demonstrated retrodirective microwave transmission using a 112-element array, 672 W of RF output, and a 6.5 m link, reporting a peak receiver-plane power density of 172 $W/m^2$. This measurement characterizes beam delivery but cannot alone establish total R4 power without spatial integration over the receiving aperture, or R6 output without electrical measurements.

Flight demonstrations require similar qualification. MAPLE's flexible coherent arrays demonstrated on-orbit WET and beamforming, including detection of an Earth-directed signal [13], whereas PRAM investigated photovoltaic-to-RF conversion without a remote receiver [38]. Both provide evidence for specific portions of the transfer chain, but neither establishes a complete measured R0-

to-R7 useful-load efficiency. Experimental validation should therefore be assigned to the energy segment actually demonstrated.

The direct-link reference planes provide a basis for interpreting reported performance across WET modalities. Section 3.2 extends this accounting approach to redirected, active, and regenerative architectures, while Section 3.3 addresses multi-node energy delivery.

### 3.2 Redirected, Active, and Regenerative Architectures

An independently located intermediate node introduces two additional reference planes: R2, the intercepted electromagnetic power, and R3, the power launched into the designated onward transfer mode. The energy pathway becomes

R0→R1→R2→R3→R4→R5→R6→R7

For passive redirection without external energy injection or release of stored energy, conservation requires

$$P_{R3} \leq P_{R2} \tag{42}$$

when both powers are evaluated over consistent boundaries and operating intervals. The difference includes dissipation and power scattered or coupled into undesired modes. Morita et al. [9] investigated untethered magnetic-resonant relay arrays and reported up to approximately 10% higher simulated transfer efficiency than linear routing over 1.5 m, illustrating how passive redistribution can improve power delivery without electrical regeneration.

A reconfigurable aperture does not necessarily constitute an intermediate relay. Li et al. [8] integrated a near-field feeder with a 13 × 13 element, 2 bit transmissive metasurface to form a compact 5.8 GHz power router. As both components belong to the transmitting assembly, the metasurface is assigned to the source-side launching function rather than a separate R2-to-R3 stage. The reported best RF-to-RF efficiency of approximately 3% [8] characterizes the overall wireless transfer system, not the metasurface alone.

Active redirection introduces electrical power to control or amplify the outgoing field. Such inputs must be recorded at the intermediate node and included in architecture-level energy accounting. Eidaks et al. [10] demonstrated a harvest-powered active relay in which the received RF signal was divided between an RF–DC harvesting branch and an RF amplification branch. Harvested energy was conditioned and stored to supply the amplifier, enabling subsequent signal forwarding. This configuration combines energy harvesting with active relaying but does not constitute complete RF-to-DC-to-RF regeneration of the forwarded power. Consequently, receiver-power enhancement alone cannot establish an end-to-end efficiency improvement without accounting for auxiliary consumption and changes in stored energy.

A fully regenerative power relay instead converts intercepted electromagnetic energy into electrical power before generating a new electromagnetic transmission. The intermediate electrical stage may incorporate power conditioning and storage, allowing successive hops to operate with different carrier frequencies, power levels, or transmission schedules. These capabilities introduce additional conversion, storage, and retransmission losses that must be included in the architecture-level energy balance. Regenerative relaying is therefore treated here as a distinct functional architecture; evidence from active RF amplification is not taken as validation of the complete regeneration pathway.

Figure 4 distinguishes these architectures according to their energy pathways, auxiliary inputs, and intermediate conversion stages. The R0–R7 labels identify the applicable energy-reference planes along each pathway, while the arrows distinguish propagating electromagnetic energy, electrical power flow, passive redirection, regenerated electromagnetic transmission, and auxiliary power. Transmitter-integrated reconfigurable apertures remain within the source-side launching function, whereas independently located passive or active redirectors introduce intermediate interception and reradiation planes. The same accounting principles extend to multi-node systems, where node availability, storage, and transfer scheduling determine useful energy delivery over time, as discussed in Section 3.3.

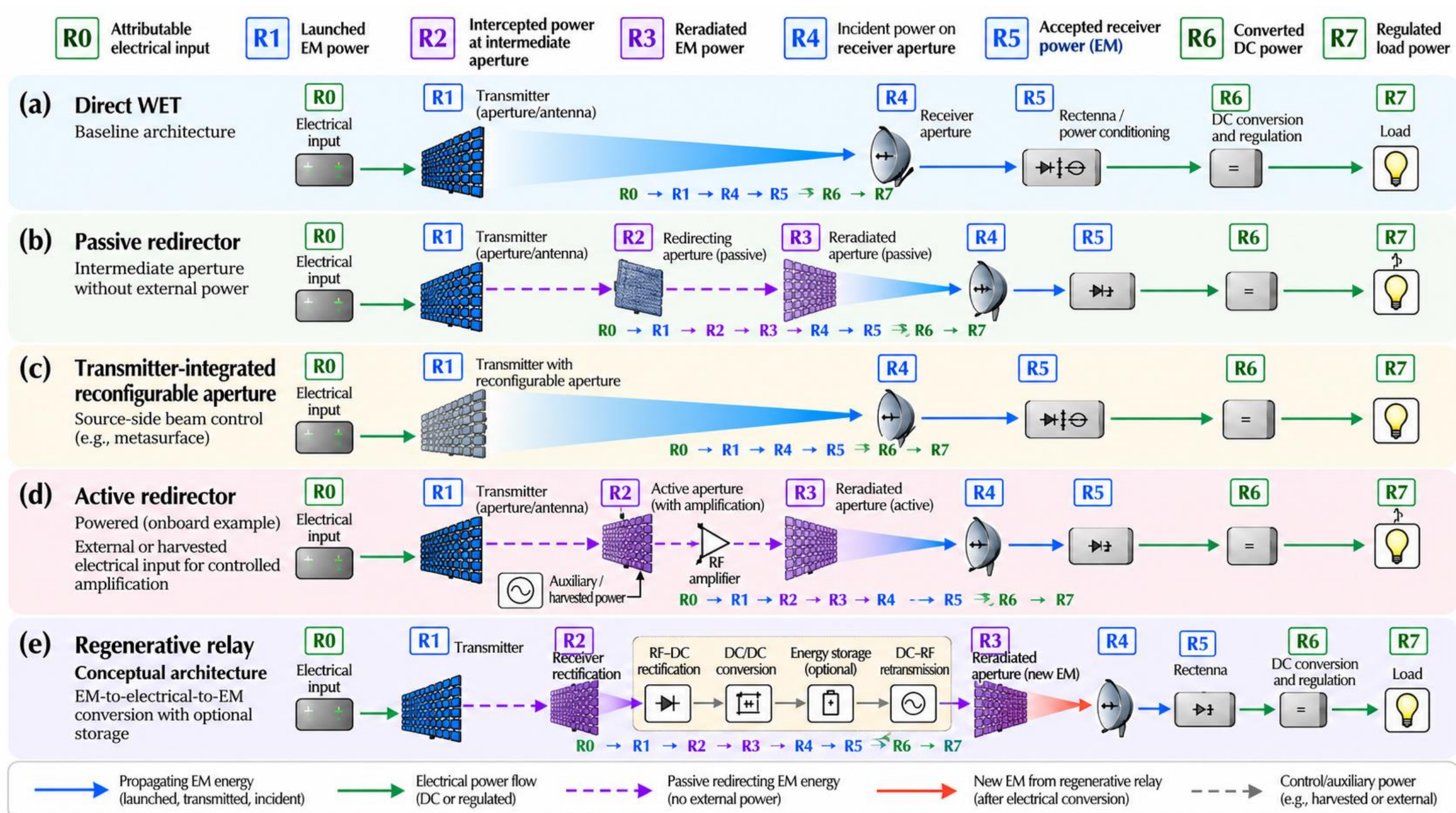

**Figure 4. Energy reference planes and functional architectures for wireless energy transfer.** R0 and R1 denote architecture-level electrical input and launched electromagnetic power; R2 and R3 identify intermediate interception and onward electromagnetic transmission; R4 and R5 represent incident power at the receiver aperture and accepted receiver power; and R6 and R7 denote converted electrical output and useful regulated load power, respectively. (a) Direct transfer without an intermediate routing stage. (b) Passive redirection through an independently located aperture without external electrical power. (c) Transmitter-integrated reconfigurable aperture, treated as part of the source-side launching function rather than as an independent intermediate node. (d) Active redirection with auxiliary electrical input, including externally supplied or harvested power for controlled RF amplification. (e) Conceptual regenerative relay incorporating electromagnetic-to-electrical conversion, optional conditioning and storage, and subsequent generation of a new electromagnetic transmission. Arrow colors distinguish propagating electromagnetic energy, electrical power flow, passive redirection, regenerated electromagnetic transmission, and control or auxiliary power. The regenerative relay is a conceptual functional architecture, not a flight-validated system.

### 3.3 Multi-Node Energy Architectures

Multi-node WET extends the architectures of Sections 3.1 and 3.2 into a time-dependent energy network. Individual links retain their underlying transfer mechanisms and reference-plane boundaries, whereas collective performance depends on source availability, geometric access, receiver demand, storage state, and scheduling. A highly efficient link may contribute little useful energy if its operating window is short or its receiver cannot accept the available power.

Storage and regenerative nodes can couple successive transfers through intermediate electrical-energy states, while constellation-level allocation coordinates multiple direct links without requiring physical relaying. Mission-level assessment must therefore consider both conversion losses and temporal availability. Within the R0–R7 accounting framework, the useful energy delivered to the final load over an interval $T$ is

$$E_{\mathrm{load}}^{(R7)}(T) = \int_0^T P_{R7}\,(t)\,dt \tag{43}$$

where $P_{R7}$ denotes useful regulated load power at reference plane R7 and T is the integration interval. These metric captures delivered power and operating duration; any corresponding mission-level efficiency requires consistent integration of attributable inputs and explicit treatment of changes in stored energy.

Alhassan et al. [70] investigated constellation-level power allocation using a simulated 20-satellite Walker 4 × 5 constellation at 450 km, transmitting 2.45 GHz microwave power to eight ground stations over 24 h. Their model incorporates orbital visibility, eclipse, battery dynamics, beam propagation, rectenna conversion, and resource allocation. Under a greedy allocation policy, the peak constellation-wide delivered DC power reached 1.986 MW, while the six served stations received mean powers of 40–75 kW each; two stations received no service [70]. The uneven delivery reflects the policy's preference for favorable instantaneous links, illustrating that link efficiency alone does not determine network-level service. These results are simulation-based and do not constitute an experimental demonstration of a space-WET network.

Multi-node design must consequently coordinate source availability, energy storage, and allocation to meet receiver demand over the mission timeline. The relevant performance measure is useful energy delivery under specified temporal and geometric constraints, rather than isolated-link efficiency. Sections 4–8 examine how these requirements affect architecture selection and implementation across space missions.

## 4. Space-to-Earth Wireless Energy Transfer

Space-to-Earth WET extends orbital energy collection into terrestrial power delivery. Building on Glaser's space solar-power concept [11], SBSP integrates orbital generation, electromagnetic transmission, ground reception, and electrical conditioning. Figure 5 compares the microwave/RF and optical/laser pathways within a common electrical transfer boundary extending from orbital photovoltaic (PV) electrical output to useful terrestrial electrical output. Microwave systems require large transmitting and receiving apertures, coherent beam control, and rectenna integration [6], whereas optical systems permit smaller diffraction-limited apertures but impose tighter requirements on pointing, atmospheric transmission, and receiver thermal management.

Sections 4.1–4.3 examine these architectures and the experimental evidence supporting their constituent functions. The central challenge is to translate demonstrated generation, transmission, reception, and conversion capabilities into sustained orbital-to-ground electrical delivery under a consistent energy-accounting framework. Section 4.4 addresses the associated beam-control, ground-reception, availability, and terrestrial-service constraints, together with the structural, thermal, operational, and economic factors that determine system-level performance.

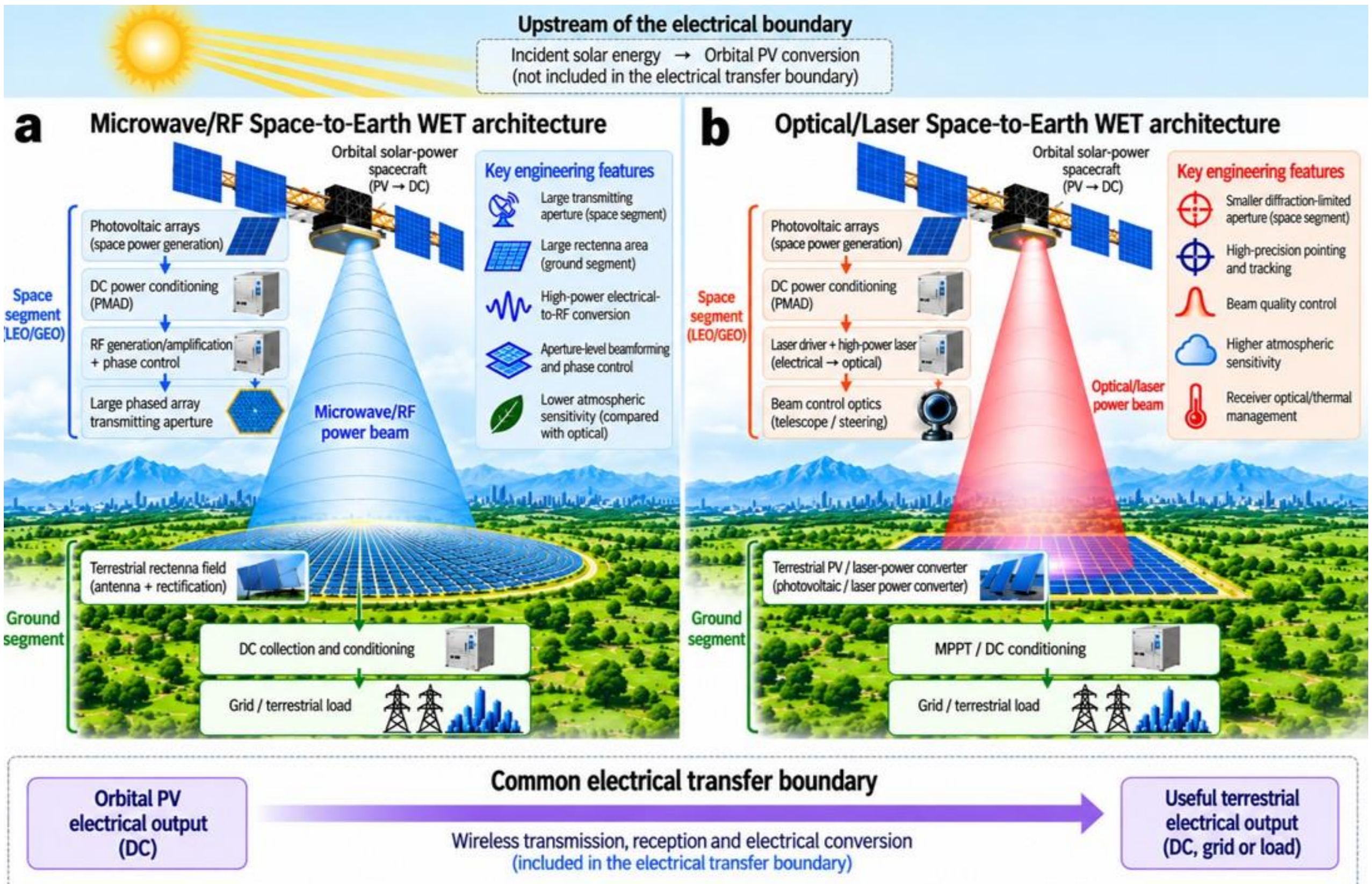

**Figure 5. Space-to-Earth wireless energy-transfer architectures within a common electrical transfer boundary.** (a) Microwave/RF architecture integrating orbital power conditioning, RF generation and coherent transmission, terrestrial rectenna reception, and downstream electrical conditioning. (b) Optical/laser architecture integrating electrical-to-optical conversion, beam control, atmospheric propagation, photovoltaic or laser-power-converter reception, and downstream electrical conditioning. The common electrical transfer boundary extends from orbital PV electrical output to useful terrestrial electrical output; incident solar energy and orbital PV conversion lie upstream of this boundary.

### 4.1 SBSP Evolution and System Architecture

Early SBSP concepts established the scale of orbital energy delivery. The NASA reference configuration examined by Arndt and Monford [71] combined orbital solar collection with a 2.45 GHz transmitting system of approximately 1 km aperture and a terrestrial rectenna designed for about 5 GW of electrical output. Although conceptual, it exposed the coupled demands of deployable structures, high-power distribution, RF conversion, phase control, thermal management, and ground reception. Subsequent designs sought to reduce the structural and integration penalties of monolithic spacecraft. NASA's Fresh Look revisited SBSP architectures in light of advances in launch systems, autonomous assembly, and power electronics [72]. CASSIOPeiA [73] emphasized modular generation and electronically steerable transmission, while ESA's SOLARIS initiative [74] addressed system concepts and critical-technology development. These studies represent architectural evolution rather than flight validation of gigawatt-scale delivery. SBSP development therefore requires coordinated advances in generation, electrical distribution, aperture stability, conversion, and thermal control within spacecraft mass and resource constraints. Microwave and optical architectures address these requirements through different transmitter and receiver configurations.

### 4.2 Microwave SBSP

Microwave SBSP has established experimental capabilities in high-power reception, beam control, and integrated conversion, although their orbital integration remains unverified. Figure 6 compares representative terrestrial microwave WPT demonstrations in terms of the performance boundaries they reveal, including receiver conversion, beam-control sensitivity, energy-boundary dependence, and practical link degradation. Because the reported metrics correspond to different measurement boundaries, they should not be interpreted as directly comparable measures of overall system performance. Dickinson's Goldstone experiment [12] delivered 30.4 kW of DC power through a 24 $m^2$ rectenna over a 1.54 km terrestrial link at 2.388 GHz, demonstrating high-power reception and rectification. A subsequent 5.8 GHz JAXA experiment [37] demonstrated phased-array transmission with electronic phase compensation for panel displacement. JAXA reported approximately 320–340 W of receiving power, with the receiving subsystem feeding a 50 V DC output before inversion to 100 V AC. Within the R0–R7 framework adopted here, this reported receiving power is therefore treated as receiver-side electrical output associated with R6, rather than as incident or intercepted microwave power at R4 or R5.

Duan et al. [75] investigated a more integrated configuration through the OMEGA prototype, combining solar-energy conversion, electrical conditioning, 5.8 GHz transmission, and rectification over 55 m. At 2081 W of transmitted microwave power, the system achieved 87.3% beam-collection efficiency and 15.05% DC–DC wireless transmission efficiency, whereas its overall solar-to-electrical efficiency was 1.15%. As illustrated in Figure 6(c), these metrics correspond to different energy boundaries and demonstrate that effective beam interception does not guarantee comparable electrical or full-system performance. Longer-distance experiments reveal additional constraints. Choi et al. [76] employed a 2.08 GHz, 2.3 kW transmitter and a 35 m antenna for approximately 1.8 km ground-to-air and ground-to-ground links. The measured aerostat power density was 154.6 $mW/m^2$, approximately 6.2% of the theoretical maximum under the study definition, rather than an end-to-end conversion efficiency. The authors identified near-field interference, antenna detuning, alignment, and rectifier mismatch as contributing factors; notably, the experimental rectifier input was substantially below its design range. Collectively, the demonstrations summarized in Figure 6 show that beam formation and interception alone do not determine useful electrical delivery; measurement boundary, receiver operating conditions, beam stability, conversion losses across the declared energy boundary, and practical link degradation must be considered jointly.

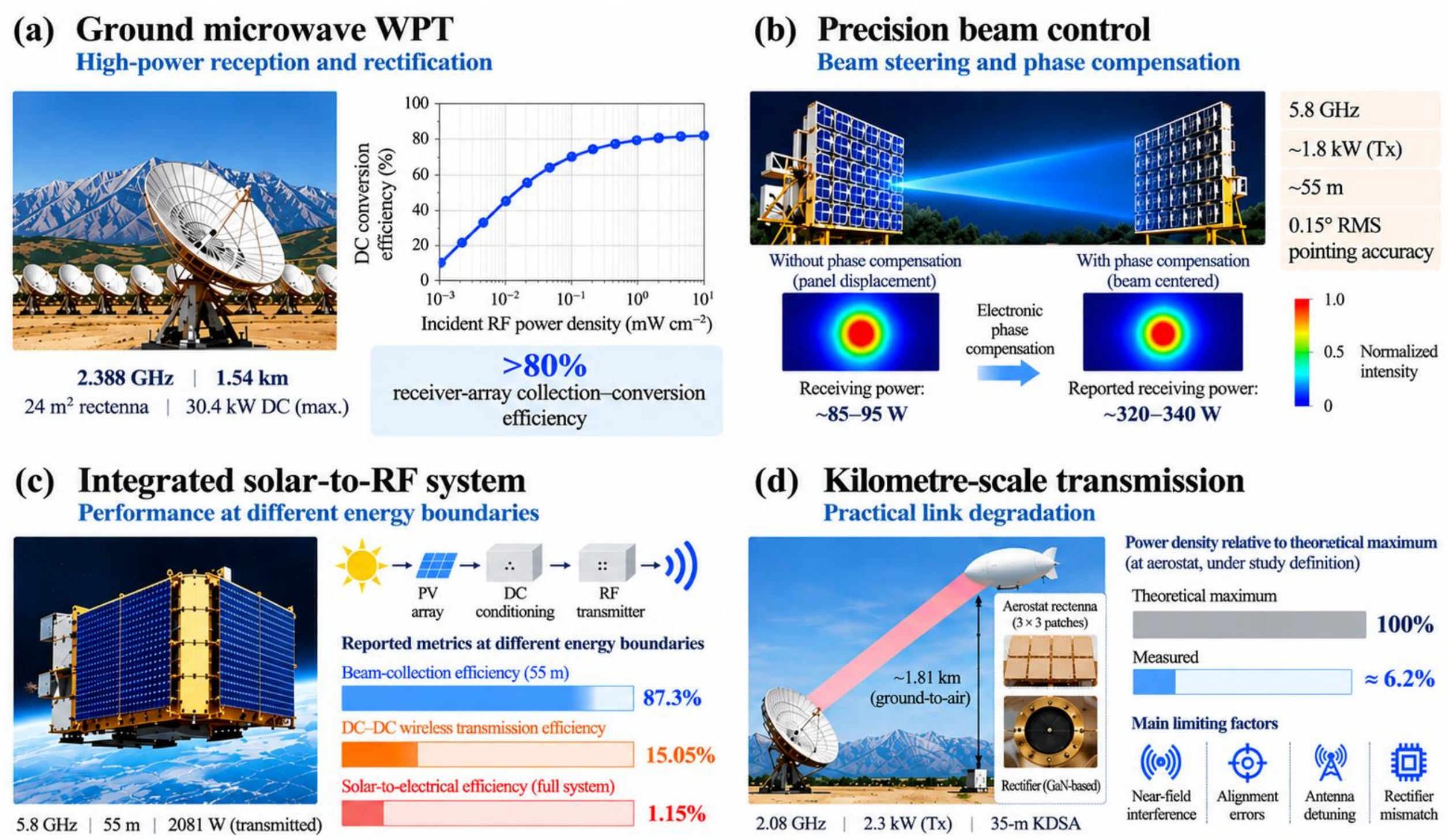

**Figure 6. Performance boundaries revealed by representative terrestrial microwave wireless-power-transmission demonstrations relevant to space-based solar power.** (a) High-power reception and rectification in the Goldstone experiment, which delivered up to 30.4 kW of DC power through a 24 m² rectenna over a 1.54 km link at 2.388 GHz, with a derived receiver-array collection–conversion efficiency exceeding 80% [12]. (b) Precision microwave beam control in the JAXA 5.8 GHz ground demonstration, showing the effect of electronic phase compensation for antenna-panel displacement on the reported receiving power [37]. (c) Integrated solar-to-RF wireless-power transmission in the OMEGA prototype, illustrating the different performance levels obtained at beam-collection, DC–DC wireless-transmission, and full solar-to-electrical energy boundaries [75]. (d) Kilometre-scale microwave transmission using a 2.08 GHz deep-space antenna, for which the measured power density at the aerostat receiver was approximately 6.2% of the theoretical maximum under the study definition, highlighting practical degradation associated with near-field interference, alignment, antenna detuning, and rectifier mismatch [76]. Reported metrics correspond to different measurement boundaries and are therefore not directly comparable across demonstrations.

### 4.3 Laser SBSP

Laser SBSP permits narrower beams from smaller physical apertures than microwave transmission, but imposes tighter requirements on pointing, beam quality, atmospheric propagation, and receiver operation. As illustrated in Fig.7(a), Ochiai et al. [63] demonstrated optical WET over 1 km under atmospheric turbulence, producing approximately 152 W of receiver electrical output from 1035 W of transmitted laser power. The reported efficiency of about 15% corresponds to the transmitted-optical-to-receiver-electrical boundary and excludes the electrical input required for laser generation.Beam shaping and receiver illumination management helped mitigate turbulence-induced nonuniformity. As shown in Fig. 7(b), DARPA's PRAD campaign [77] extended terrestrial optical power beaming to 8.6 km, delivering more than 800 W during a 30s interval and transferring over 1 MJ across the broader campaign. At shorter distances, optical-output-to-receiver-electrical conversion exceeded 20%. Together, the terrestrial demonstrations summarized in Fig.7 extend the demonstrated power–distance range but do not establish orbital acquisition, long-duration availability, spacecraft thermal management, or electrical source efficiency. Laser SBSP must therefore integrate efficient optical generation, stable propagation, photovoltaic conversion, and heat rejection under changing orbital and atmospheric conditions.

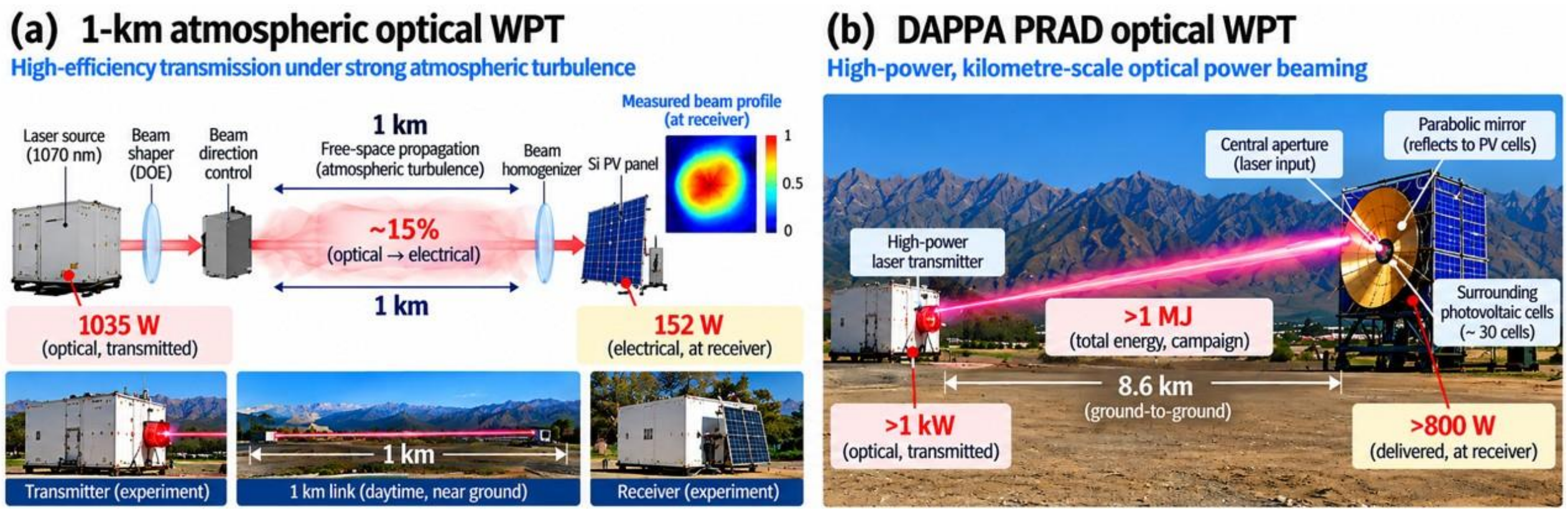

**Figure 7. Experimental evidence for long-distance optical wireless power transfer relevant to laser SBSP.** (a) Kilometre-scale atmospheric optical WPT under strong turbulence, incorporating transmitter-side beam shaping and receiver-side homogenization for photovoltaic power extraction; 1035 W of transmitted optical power produced approximately 152 W of receiver electrical output over a 1 km link, corresponding to a reported optical-to-electrical efficiency of approximately 15% within the stated experimental boundary [63]. (b) DARPA's PRAD demonstration of high-power optical power beaming over an 8.6 km terrestrial link, delivering more than 800 W at the receiver during a 30 s transmission and transferring more than 1 MJ over the broader experimental campaign [77]. These demonstrations establish kilometre-scale terrestrial atmospheric propagation and high-power receiver delivery but do not constitute end-to-end orbital laser-SBSP validation.

### 4.4 Beam Control, Ground Receiving Systems, and Terrestrial Service

Large SBSP apertures are sensitive to deployment errors, thermoelastic deformation, vibration, and attitude motion, which perturb the transmitted wavefront and receiver interception. MAPLE demonstrated coherent beamforming with flexible arrays in low Earth orbit, including detection of an Earth-directed signal at a ground station [13], providing flight evidence for lightweight aperture control. Orbital geometry also determines energy availability. Peters et al. [78] modeled a 10,000 km medium-Earth-orbit configuration at 55° inclination, reporting 28% overhead time, 30% Earth coverage, and 97% solar capture under their assumptions. Design refinements increased modeled transmission efficiency by approximately 5–10%. Complementary moving-platform evidence comes from the Japan Space Systems JFY2023/2024 program [79], which demonstrated 5.8 GHz phased-array transmission from an aircraft to a ground receiver over more than 5 km. Terrestrial service requires more than electromagnetic interception. Microwave and optical receivers must convert and condition collected power before grid or load delivery, while availability, protection, and operational safety determine whether intermittent transmission can provide useful service. Economic feasibility introduces another constraint. NASA's 2024 assessment [80] estimatedbaseline levelized electricity costs of \$0.61/kWh and \$1.59/kWh for two representative 2 GW SBSP designs intended for operation beginning in 2050. Under the study assumptions, these values were approximately 12–31 and 32–80 times the projected costs of the terrestrial alternatives considered, respectively. Launch and manufacturing dominated the baseline estimates, although the results were sensitive to assumed technological and operational improvements. The next system-level milestone is an orbital demonstration linking electrical generation, controlled transmission, terrestrial reception, and useful load delivery within a consistent energy budget. Mass, thermal performance, pointing stability, and availability must be assessed alongside electrical output. Scaling SBSP from an individual transmission system to a constellation introduces an additional constraint that is not captured by link efficiency or single-station power alone: the number of orbital and terrestrial locations at which complete source-to-receiver pathways can actually be realized. Atzori et al. [81] evaluated the intrinsic orbital capacity of GEO SBSP constellations under progressively stricter assumptions involving minimum station separation, the existing GEO satellite population, terrestrial receiver geography, and population-density-based siting constraints. For a minimum distance angle of 0.1°, their modeled capacity decreased from an idealized 3600 stations to 364 available slots in the most restrictive scenario. Coupling this deployment model to a time-dependent solar-generation model yielded annual grid-energy estimates ranging from approximately 8903 to 900 TWh $yr^{-1}$ across the four scenarios. These values are scenario-dependent model outputs rather than demonstrated SBSP performance, but they illustrate that large-scale energy delivery depends not only on the performance of an individual transmission chain, but also on orbital availability and the realizability of the associated ground segment.

Thus, scaling SBSP from a single transmission chain to an energy infrastructure introduces a capacity constraint outside the conventional link-efficiency boundary: useful system scale is determined not only by how efficiently one source–receiver pair transfers power, but also by how many complete orbital–terrestrial pathways can be deployed under the relevant orbital and ground-segment constraints. Beyond terrestrial service, however, WET faces a different operating problem in lunar and cislunar environments, where the principal requirement may be to deliver energy across terrain, mobility, and illumination constraints rather than to maximize continuous grid-scale output.

## 5. Cislunar and Lunar Wireless Energy Infrastructure

Lunar WET addresses the spatial and temporal separation between energy resources and surface demand. Permanently shadowed regions (PSRs), prolonged darkness, and terrain-dependent illumination constrain local solar generation, while exploration, habitation, and in situ resource utilization (ISRU) impose increasingly diverse power requirements [1, 2]. NASA identifies demands ranging from approximately 2 kW for selected uncrewed configurations to 20 kW for crewed habitats and at least 60 kW for certain ISRU applications [82].Colozza [83] separately estimated 25.83 kW for an oxygen-production facility and 28.05 kW for a six-person base. These requirements provide a reference for evaluating WET, rather than representing demonstrated wireless-delivery capabilities. The role of WET must be assessed within the complete lunar energy system, including local generation, storage, and distribution [84].

As illustrated in Fig. 8, orbit-to-surface transfer supplies lunar receivers from orbital platforms, whereas surface-to-surface transfer redistributes locally available energy across terrain. Multi-orbiter and hybrid architectures extend these pathways toward coordinated energy service by combining orbital WET with local generation, storage, and load management. These configurations address different mission constraints rather than successive stages of a single technology. Their applicability depends on useful electrical output, access duration, environmental conditions, receiver conversion, and storage requirements. Figure 9 complements this architectural overview by comparing representative constraints on power and energy delivery, orbital access and storage, and surface-link performance.

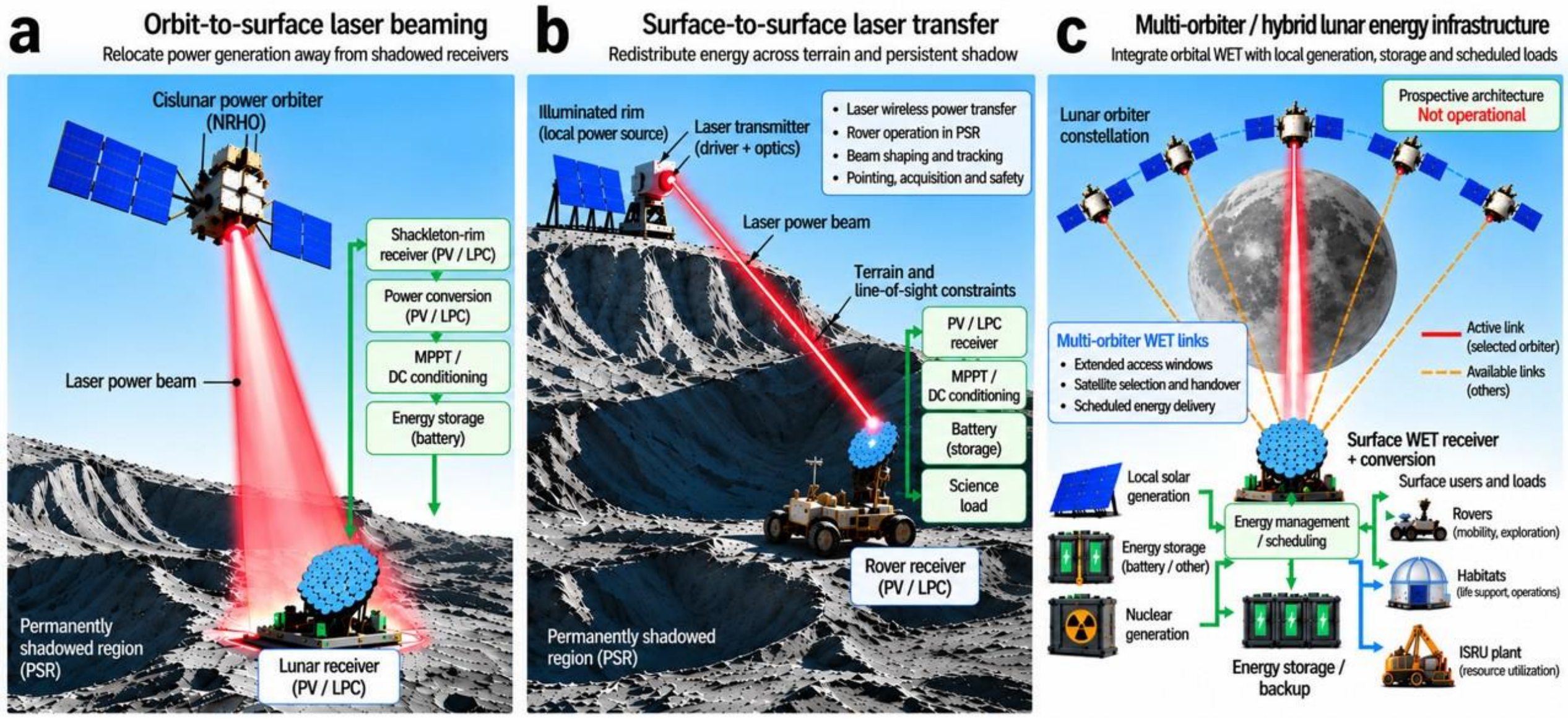

**Figure 8. Wireless energy transfer architectures for lunar and cislunar missions.** (a) Orbit-to-surface laser power delivery from a cislunar platform to a lunar receiver, illustrating receiver conversion, power conditioning, and storage [84]. (b) Surface-to-surface laser transfer from an illuminated site to a shadowed or mobile receiver, highlighting terrain, line-of-sight, beam-control, and receiver constraints [85]. (c) Prospective multi-orbiter and hybrid lunar energy infrastructure integrating orbital WET with local generation, storage, energy management, and heterogeneous surface loads [83,86–88]. The panels represent alternative architectural pathways; the multi-orbiter/hybrid configuration remains prospective.

### 5.1. Mission-Driven Power Delivery: Orbital and Surface Links

Orbit-to-surface WET can supply terrain-constrained receivers without continuous surface transmission infrastructure, but its performance depends on orbital visibility, propagation distance, and receiver conversion. Early rover and science-lander studies established this approach [35, 89, 90]. Lopez et al. [88] subsequently reported maximum converted powers of 18.92 and 21.61 kW for representative ten-satellite circular-polar and distant retrograde-orbit configurations, respectively. These modeled peak values differ from sustained energy delivery. Turyshev [84], for example, predicted approximately 0.66 kWh/day, equivalent to an average of 28 W, for a reference near-rectilinear halo-orbit system serving the Shackleton rim. The associated parametric analysis further illustrates the dependence of daily energy delivery on transmitted power and transmitter aperture under the specified assumptions (Fig. 9(a)). The peak-power and daily-energy results therefore describe different configurations and performance boundaries and should not be directly combined. Orbital WET must be evaluated through both instantaneous output and energy delivered over the available access windows.

NASA's OPAL study [91] links orbital access to client power and storage requirements. For a modeled 200 W lander with a 1.3 kWh battery, increasing the constellation from one to three spacecraft reduces the approximate battery-only operating interval from about 3 h to about 20 min under the selected assumptions (Fig.9(b)). The corresponding state-of-charge histories illustrate the coupling between access continuity and surface-storage requirements. More frequent access may reduce the required storage burden, although this benefit must be weighed against additional spacecraft, generation, and thermal-control hardware. Sfasciamuro et al. [92] further investigated orbital laser-system integration and measured photonic power-converter efficiencies of 14.6–16.3% under laboratory conditions. These receiver-level results do not establish end-to-end lunar delivery performance.

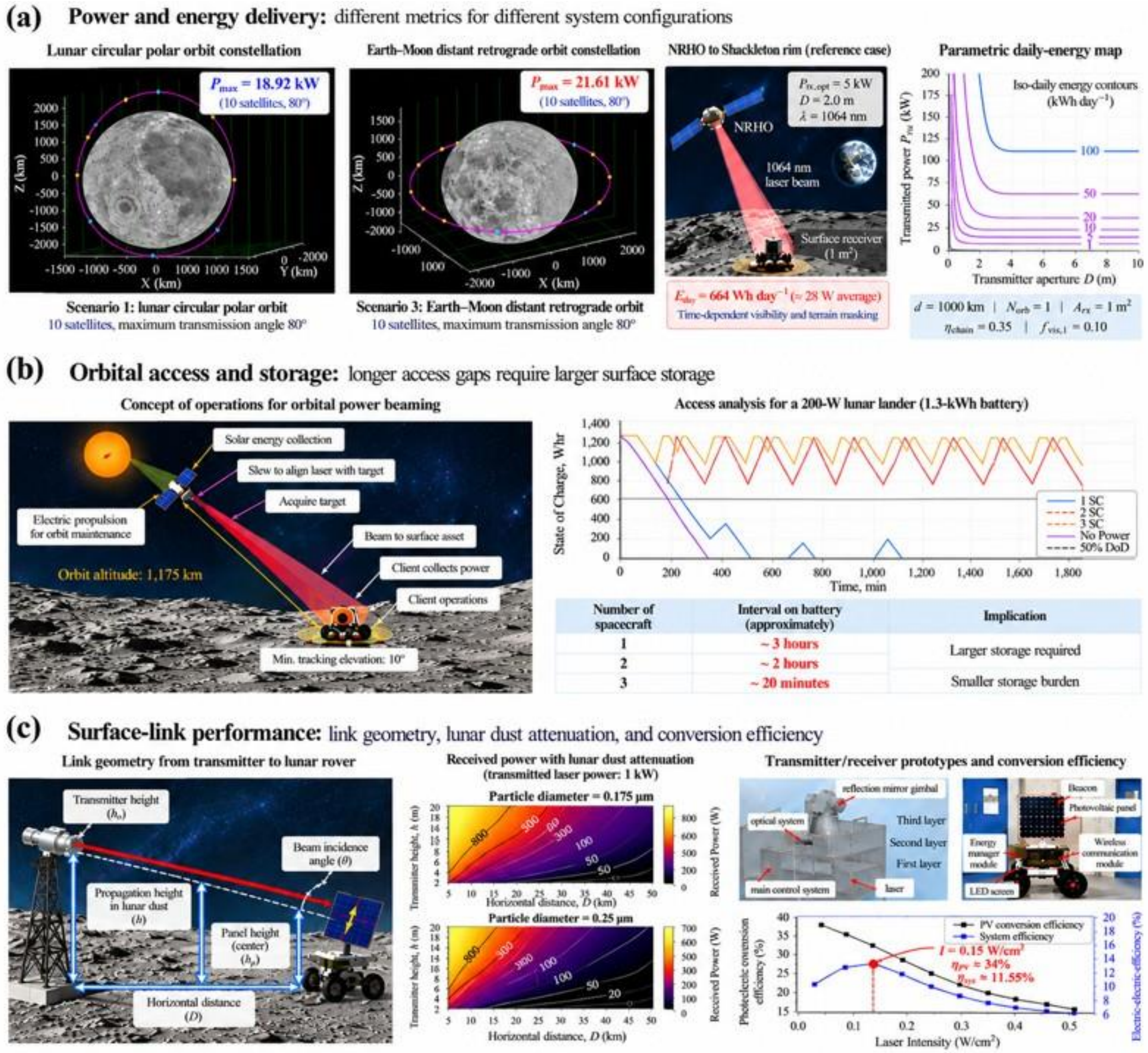

**Figure 9. Mission-performance constraints for lunar wireless energy transfer.** (a) Representative orbital results distinguish modeled maximum converted power for ten-satellite circular-polar and distant-retrograde-orbit configurations [90] from time-integrated energy delivery for a separate NRHO-to-Shackleton reference case [84]. The parametric map illustrates daily-energy dependence on transmitted power and transmitter aperture under the assumptions of [84]. (b) OPAL concept of operations and modeled state-of-charge response for a 200-W lunar lander with a 1.3kWh battery, illustrating the coupling between orbital access and surface storage [91]. (c) Surface-link geometry and modeled dust-dependent received power [95], together with terrestrial transmitter/receiver prototypes and conversion characteristics [85, 87]. The panels distinguish modeled mission performance, terrestrial experimental evidence, and different system boundaries rather than representing a single end-to-end demonstration.

Surface-to-surface WET offers a complementary route from illuminated ridges or elevated stations to shadowed and mobile receivers. It avoids orbital access interruptions but requires suitable terrain clearance, transmitter placement, and receiver tracking. Xu et al. [85] demonstrated 11.55% electrical-to-electrical efficiency over 10 m on Earth; the measured photovoltaic and system-efficiency trends further show the dependence of conversion performance on incident laser intensity (Fig.9(c)). Tian et al. [93] formulated PSR supply as a coverage–connectivity–cost problem, while the proposed BEACON node [94] combines energy delivery and communications, targeting up to 500 W per user over distances approaching 10 km; these values remain design objectives rather than lunar demonstration results.

Environmental uncertainty becomes increasingly important for extended surface links. Jiwan-Mercier et al. [95] modeled substantial variations in received power with distance, source height, and dust assumptions. As illustrated in Fig.9(c), the predicted link performance changes markedly with propagation distance and the assumed 0.175 and 0.25μm dust-particle cases. Tian et al. [96] similarly identified strong sensitivity to beam elevation while acknowledging the absence of direct dust measurements in polar PSRs. Accordingly, predicted attenuation should be treated as a scenario-dependent design input rather than a universal lunar loss factor. The terrestrial prototype and lunar propagation models in Fig. 9(c) therefore represent complementary but distinct evidence boundaries.

The two pathways serve different operating conditions. Orbital systems can bypass surface obstacles but require management of access windows; surface links exploit nearby generation but depend on terrain and environmental exposure. Intermittent charging may support rovers and distributed instruments, whereas continuous ISRU and habitation loads require more demanding provisions for generation, storage, and redundancy. Figure 9 shows that peak power, time-integrated energy, access continuity, propagation conditions, and receiver conversion provide complementary measures of mission performance. Mission suitability must therefore be determined by useful energy delivered over the required operating period, together with the burden of the complete power system.

### 5.2. Multi-Node Energy Service and Hybrid Lunar Power Systems

Extending individual WET links into energy infrastructure requires coordinated source selection, storage, and load management. Figure 8(c) illustrates this transition toward a prospective multi-node architecture integrating orbital WET with local generation, storage, energy management, and heterogeneous surface loads. Constellation studies by Lopez et al. [88] and Dönmez et al. [86] examine orbital coverage and handover, while OPAL [91] relates supply frequency to client storage. Additional transmitters can shorten interruptions, but also increase spacecraft mass, generation requirements, thermal-control demands, and operational complexity. Geometric coverage alone does not guarantee electrical service: receiver acquisition, conversion, and energy availability must coincide with load demand.

Surface networks face similar trade-offs. Elevated or distributed nodes may extend coverage across obstructed terrain [93, 94], although shared infrastructure introduces common failure points. WET should therefore complement rather than automatically replace conventional distribution. NASA's LunaGrid-Lite project [97] is developing a cable-based demonstration concept for approximately 1 kW over 100–500 m, while Gao et al. [1] examine wired and wireless transmission within broader lunar grid architectures. Physical connections may suit accessible fixed installations, whereas non-contact delivery offers flexibility for mobile or geographically isolated users. Hybrid systems can combine local photovoltaics, nuclear generation, storage, wired distribution, and WET according to mission requirements.

Ijichi et al. [87] examined such an integrated lunar power-system concept. Within this arrangement, local sources can support persistent loads, storage can bridge interruptions, and wireless links can deliver energy to remote receivers. Power-management and distribution hardware must regulate variable receiver output, coordinate charging, and maintain load continuity during source switching. The principal limitation remains the gap between subsystem evidence and integrated service. Terrestrial experiments have demonstrated short-range electrical transfer [85], and laboratory studies have characterized components for orbital architectures [92]; constellation and grid performance, however, remains largely model-based [86–88, 91]. Future demonstrations should measure transmitter electrical input, emitted power, receiver electrical output, and regulated load energy over representative access cycles, together with pointing, thermal, and environmental performance. For multi-node systems, source availability, link handover, storage state of charge, and load continuity should also be resolved over the same operating interval. The value of lunar WET will ultimately depend on whether these systems can deliver the required energy with an acceptable combined burden of generation, transmission, storage, and operation.

## 6. Spacecraft-to-Spacecraft and Distributed Orbital Energy Sharing

Spacecraft-to-spacecraft WET offers a means of reallocating power-generation and storage capacity among cooperating vehicles. A dedicated provider could supply clients equipped with receiving apertures, conversion electronics, and residual storage, avoiding the need to size every spacecraft for its peak demand. Whether this arrangement reduces system mass, however, depends on the additional transmission hardware, pointing requirements, service availability, and redundancy. Unlike a conventional spacecraft power system, shared generation also introduces dependence on the provider and the orbital geometry.

Early concepts established several possible configurations. Power Plug proposed an approximately 100 kWe-class provider using RF or optical transmission [3], while Jamnejad and Silva [98] investigated microwave delivery from a mother spacecraft to moving mini-spacecraft. Larger networks have been explored by Asakawa et al. [99], who considered 24 laser-transmission satellites serving two groups of 24 clients at 300 and 1800 km, and Baraskar et al. [100], whose Energy Orbit concept comprised 1600 power satellites in LEO. These studies define alternative service architectures, although their projected mass savings and constellation performance remain unverified.

Experimental progress has addressed individual elements of this concept. Yuasa et al. [101] demonstrated a 24 GHz phased-array transceiver integrating local-oscillator transfer and dual-circular-polarized WPT, obtaining 3.8 mW in a ground experiment. The RINGS experiment subsequently integrated resonant-inductive transfer with SPHERES hardware aboard the ISS and reported approximately 30% efficiency in preliminary WPT testing [35]. Separate formation-flight tests involved restrained and free-floating units, but the reported efficiency does not establish sustained transfer during relative motion. RINGS therefore provides short-range microgravity evidence rather than validation of a radiative orbital service.

Radiative technologies have reached a different level of maturity. MAPLE demonstrated flexible phased-array operation, wireless transmission, and Earth-directed signal detection in orbit, while revealing thermal and power-output degradation that remains relevant to long-duration operation [13]. The proposed 16U4SBSP mission extends the intended scale to a seven-CubeSat, kilowatt-class RF demonstration [102]. OHISAMA instead targets a separated transmitter–receiver configuration with a released probe operating several meters to approximately 50 m from the source [103]. Its terrestrial precursor delivered 12 W to a rectenna at 5 m under equal-amplitude, equal-phase excitation, increasing to 24 W when the beam was focused at the receiver [103]. These results establish important hardware capabilities, but useful electrical delivery between independently flying spacecraft has yet to be demonstrated.

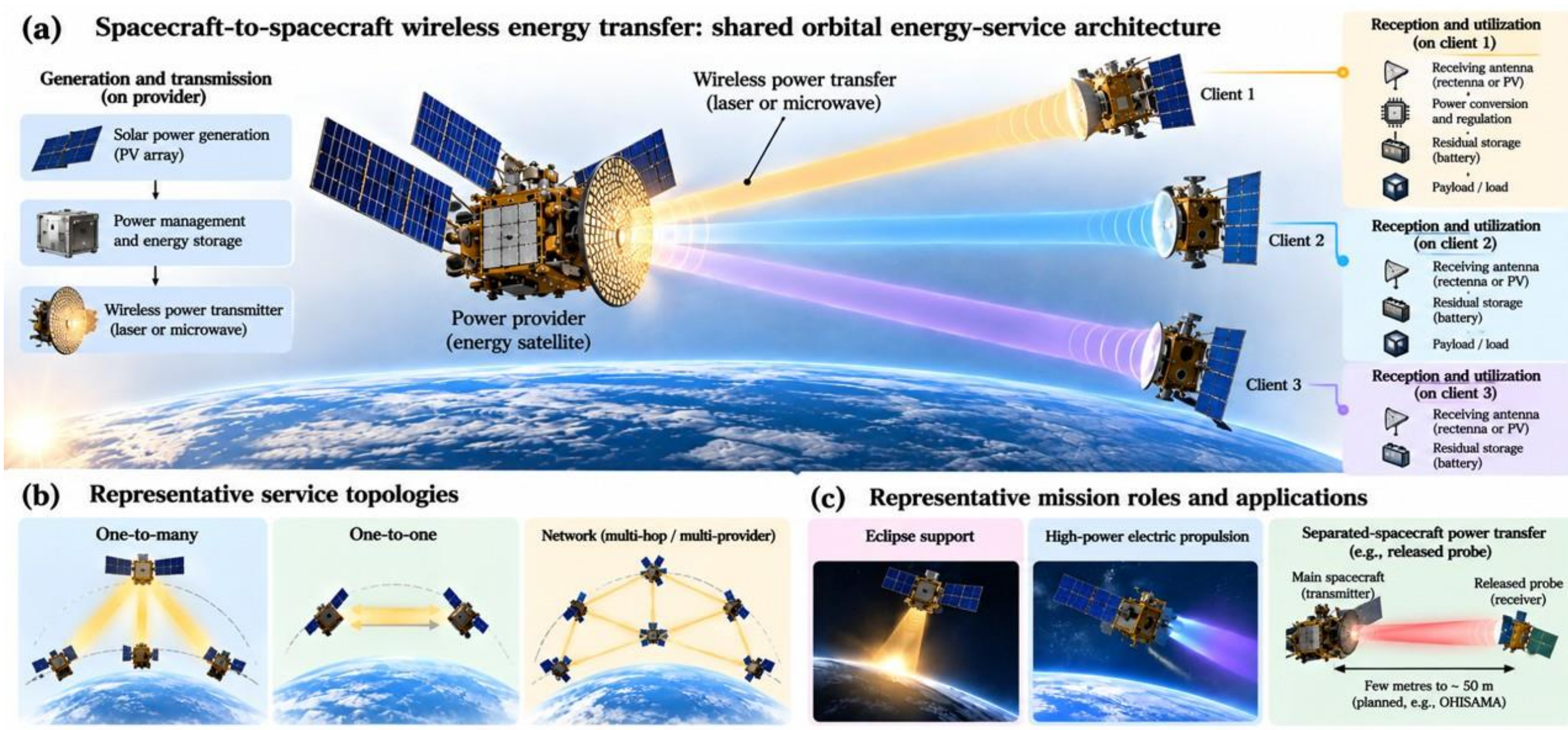

**Figure 10. Spacecraft-to-spacecraft WET as a shared orbital energy-service architecture.** (a) A dedicated provider supplies one or more clients, shifting generation and part of the storage burden toward the provider while retaining client-side reception, conversion, residual storage, and load interfacing. (b) Representative service topologies include one-to-many, one-to-one, and networked delivery, abstracting concepts in Refs. [3,98-10]. (c) Candidate mission roles include eclipse support [75], electric-propulsion support [104], and formation or constellation power sharing. The architectures are prospective system configurations rather than demonstrations of closed electrical energy service between independently flying spacecraft.

Figure 10 organizes these concepts as a shared orbital energy-service architecture. A dedicated provider may serve one or more clients through one-to-one, one-to-many, or networked configurations, redistributing generation and storage requirements rather than eliminating them. Representative mission roles include eclipse support, high-power electric propulsion, and formation or constellation power sharing.

The prospective benefit is particularly clear for eclipse support and intermittent high-power loads. Peters et al. [78] modeled RF delivery from an 11,000 km satellite to a client at 1000 km, obtaining approximately 90.9% coverage during a selected eclipse interval. Such access could reduce client storage requirements, although the associated battery-mass saving was not measured. For electric-propulsion missions, Dankanich et al. [104] proposed separating generation from the transfer vehicle. Their model increased effective specific power from 4 to 40 W/kg and reduced predicted thrusting times from 118 to 12 days for GTO-to-GEO transfer and from 235 to 24 days for LEO-to-GEO transfer, assuming a 2000 s Hall propulsion system. These gains depend on the mass and availability of the external power infrastructure.

The remaining challenge is therefore architectural rather than the feasibility of wireless transmission alone. A convincing orbital energy-service demonstration must deliver regulated electrical power to a moving client over representative acquisition, tracking, and interruption cycles, while accounting for the provider, receiver, residual storage, and failure tolerance. Only then can shared generation be assessed against conventional onboard power systems on a consistent mission basis.

## 7. Deep-Space and Extreme-Environment Applications

Deep-space WET is governed not simply by transmission distance, but by the constraints separating favorable generation sites from demanding loads. Solar availability, atmospheric attenuation, surface temperature, and power demand can alter both source location and delivery architecture. As summarized in Fig. 11, Mars, Venus, and Miranda illustrate three distinct mission-level constraints: load-dependent system benefit, atmospheric transmission and receiver conversion, and long-range optical delivery, respectively.

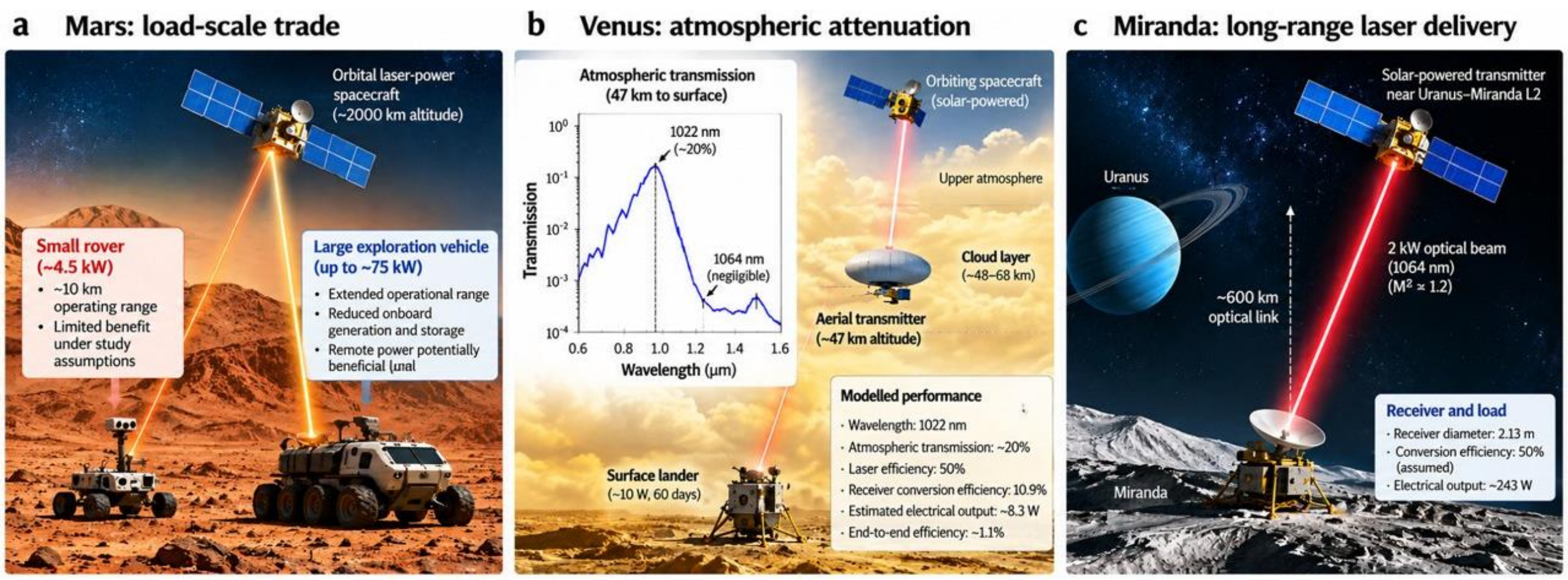

**Figure 11. Wireless energy-transfer concepts and mission-level constraints for deep-space applications.** (a) Mars: conceptual orbital laser-power delivery to exploration vehicles with different load requirements, illustrating the load-dependent value of remote generation under the modeled assumptions [4]. (b) Venus: atmospheric attenuation motivates an aerial transmitter near the cloud base (~47 km); the modeled transmission window at 1022 nm (~20%) and high-temperature receiver conversion limit the estimated lander electrical output to ~8.3 W [105,106]. (c) Miranda: long-range laser delivery from a solar-powered transmitter near the Uranus–Miranda L2 region to a surface lander over ~600 km, highlighting aperture, beam quality, pointing, receiver collection, and thermal-management requirements [107]. All results represent mission concepts or model-based estimates rather than flight-demonstrated WET performance; schematics are not to scale.

Mars highlights the importance of load scale. Harries et al. [4] compared a small unpressurized rover requiring approximately 4.5 kW over a 10 km operating range with a larger exploration vehicle demanding up to 75 kW. The latter concept employed three orbiting laser platforms at approximately 2000 km. Under the study assumptions, remote power offered limited benefit for the smaller rover but potential advantages for the larger vehicle by reducing onboard generation and storage requirements (Fig. 11a). More recent work has examined orbital delivery and alternative beam paths [108], although the evidence remains model-based.

On Venus, atmospheric attenuation influences transmitter location, while extreme surface temperatures constrain electrical conversion. A NIAC concept considered a lander requiring approximately 10 W over 60 days [106]. Grandidier et al. [105] investigated an aerial transmitter near the cloud base at approximately 47 km. Their model predicted atmospheric transmission of approximately 20% at 1022 nm but negligible transmission at 1064 nm. Assuming a 5 $m^2$ solar array, 50% laser efficiency, and 10.9% receiver conversion efficiency, the estimated lander output was approximately 8.3 W, corresponding to approximately 1.1% end-to-end beaming efficiency. As illustrated in Fig. 11b, the atmospheric transmission window therefore affects wavelength and transmitter-location selection, while receiver conversion remains a major constraint on useful electrical output. The improved propagation path remains subject to platform endurance, pointing, and thermal-management requirements.

### 7.1 Small-Body and Icy-Moon Applications

No dedicated asteroid or comet study establishing a complete electrical-WET pathway was identified in the literature examined here. Laser ablation, deflection, and beamed propulsion are excluded because they do not deliver electrical energy to a receiving load. Icy-moon applications introduce substantial demands on source capacity, optical delivery, and receiver collection.

The Miranda reference mission [107] proposes a solar-powered platform near the Uranus–Miranda L2 region transmitting 2 kW of 1064 nm laser power over approximately 600 km. With a 0.5 m transmitting aperture, a 2.13 m receiver, and an assumed 50% receiver-side conversion efficiency, the design predicts approximately 243 W of lander electrical output. As illustrated in Fig. 11c, this long-range link couple's transmitter aperture, beam quality, pointing stability, receiver collection, and conversion performance. Beam quality ($M^2$) and pointing stability are therefore critical, while the assumed 50% electrical-to-optical efficiency implies approximately 4 kW of electrical input and 2 kW of conversion waste heat, excluding other subsystem losses [107]. These values define a mission-level reference design, not demonstrated outer-Solar-System performance or transmission through subsurface ice.

Across the three cases in Fig. 11, the dominant constraint shifts from load-dependent system benefit on Mars to atmospheric transmission and high-temperature conversion on Venus, and to aperture–pointing–thermal coupling at Miranda. Their modeled outputs are not directly comparable without consistent energy and mission boundaries. The suitability of deep-space WET must ultimately be established against local solar, storage, or nuclear alternatives under equivalent power, lifetime, and operational requirements.

## 8. Intra-Spacecraft Contactless Power Transfer

Intra-spacecraft WET primarily addresses the limitations of conventional electrical interfaces rather than long-distance power delivery. Across moving, thermally isolated, or repeatedly assembled structures, contactless transfer can eliminate sliding contacts and simplify wiring. These benefits must nevertheless justify the additional coupling hardware, conversion losses, mass, and electromagnetic-compatibility requirements relative to cables, connectors, or slip rings. Unlike the inter-spacecraft energy-service architectures discussed in Section 6, the focus here is contactless power delivery across local interfaces within or directly associated with a spacecraft.

Early ground experiments by Vazquez Ramos and Yuan [109] demonstrated approximately 4 W across a 3 cm gap and powered a battery-free handheld radio at distances of up to 5 cm. For higher-power rotary applications, a NASA-supported 100 kW class, 20 kHz design study selected a rotary transformer after comparing magnetic, capacitive, and microwave alternatives [110]. Modular interfaces impose different requirements, including alignment tolerance, repeated mating, and bidirectional operation. Zhang et al. [111] demonstrated a ground-based magnetic interface transferring approximately 20–30 W per single-stage link with at least 75.8% reported efficiency. Figure 12(a) illustrates this interface-level demonstration. Although these studies establish feasible interface configurations, they do not constitute mission-lifetime qualification.

### 8.1. Sensors, Extravehicular Electronics, and Near-Field Integration

Contactless supply is particularly relevant to distributed sensors where wiring and electrical feedthroughs complicate installation. Kishikawa and Kawasaki [112] developed a 5.8 GHz satellite-sensor-oriented system whose GaN–Si hybrid rectifier achieved 10.3% RF-to-DC efficiency. The complete 0.2 m experimental link, however, reached only 0.014% conversion efficiency when antenna, cable, propagation, and rectification losses were included. Figure 12(b) illustrates this boundary dependence directly: component-level rectifier efficiency should not be conflated with complete-link efficiency. This discrepancy illustrates why receiver-component performance alone cannot establish the viability of a spacecraft power interface.

Magnetic coupling can also deliver power across thermally isolated or mechanically separated structures, although conductive surroundings, shielding, and packaging affect both transfer performance and electromagnetic compatibility. To reconcile wireless transfer with spacecraft insulation, Honda et al. [113] developed magnetic-transmissive multilayer insulation (MT-MLI). Their WPT experiments achieved 90.7% coil-to-coil efficiency, compared with 54.0% using conventional MLI. The corresponding configurations are compared in Fig.12(c). Separate thermal-vacuum tests demonstrated reduced heat leakage and reported 72% transmission efficiency during thermostat-controlled operation. These results establish the feasibility of combining thermal insulation and magnetic transfer, but not complete DC-to-load performance or flight qualification.

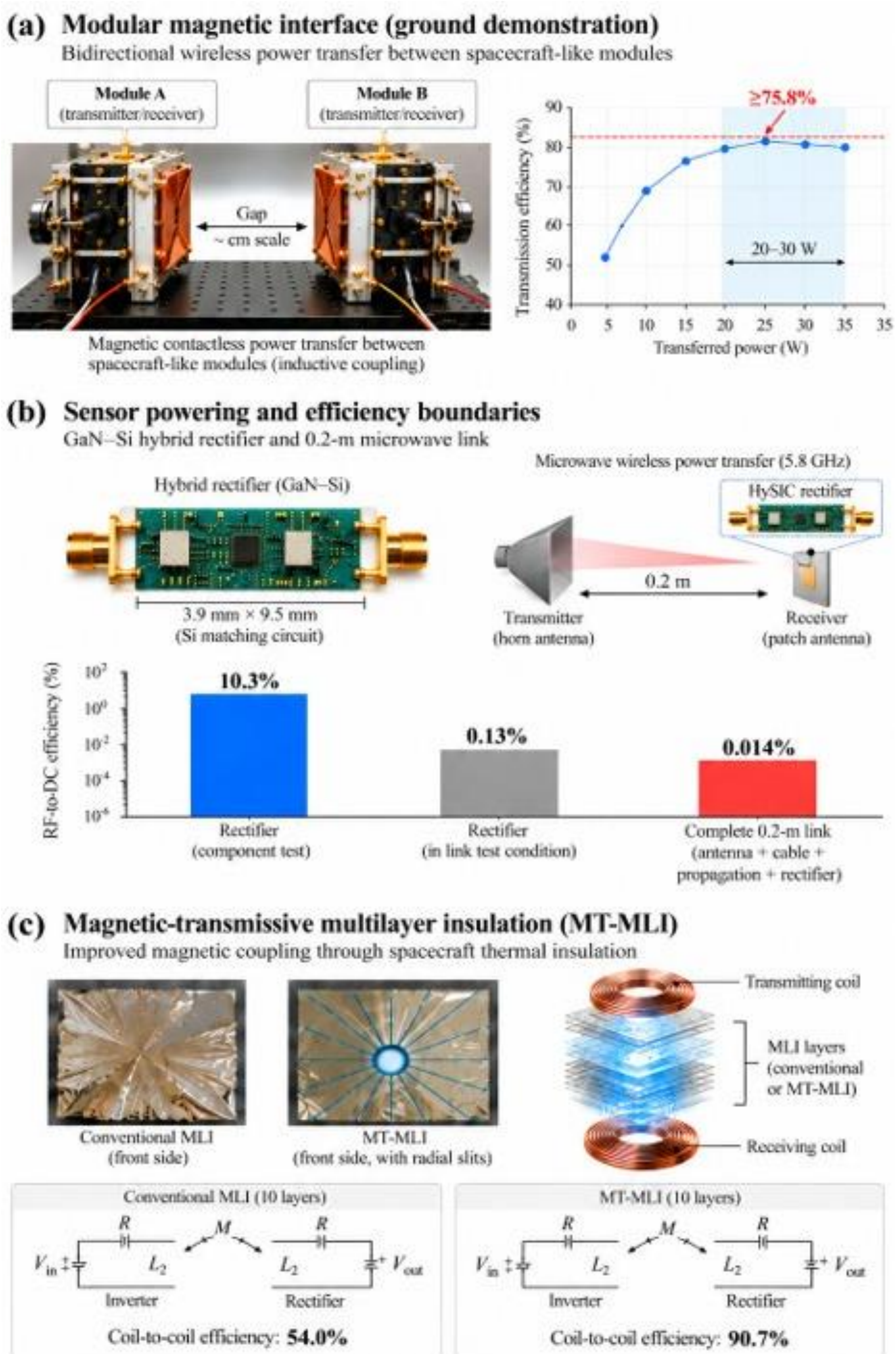

**Figure 12. Interface-level evidence for intra-spacecraft contactless power transfer.** (a) Ground-demonstrated bidirectional magnetic power transfer between modular spacecraft-like interfaces, illustrating 20–30 W operation with a reported transmission efficiency of at least 75.8% [111]. (b) Sensor-oriented microwave power transfer illustrating distinct efficiency boundaries between a GaN–Si hybrid rectifier and the complete 0.2-m experimental link [112]. (c) Integration of magnetic power transfer with spacecraft thermal insulation using magnetic-transmissive multilayer insulation (MT-MLI), which increased the reported coil-to-coil efficiency from 54.0% with conventional MLI to 90.7% [113]. The reported efficiencies correspond to different experimental boundaries and should not be interpreted as directly comparable system-level efficiencies.

### 8.2. Flight Demonstrations and Qualification Status

In-orbit experiments have validated distinct short-range WET functions (Fig. 13). LEctenna demonstrated milliwatt-scale embedded RF powering aboard the ISS [114]. The Space Wireless Energy Laser Link (SWELL), installed on the ISS STP-H9 platform, subsequently demonstrated sustained optical transfer across 1.45 m, delivering approximately 1.5 W at the photovoltaic receiver output with an NRL-reported end-to-end efficiency of approximately 11% and more than 100 days of successful operation reported by June 2023 [39]. Tianzhou-7 provided complementary evidence for integrated near-field wireless power distribution. Launched in January 2024, it carried an experiment combining magnetic shielding with mechanical, electrical, and thermal design, with a reported in-orbit wireless transmission efficiency of 91.7% [40, 41]. This value corresponds to the reported experimental transmission boundary and should not be interpreted as the overall efficiency of the spacecraft electrical power system.

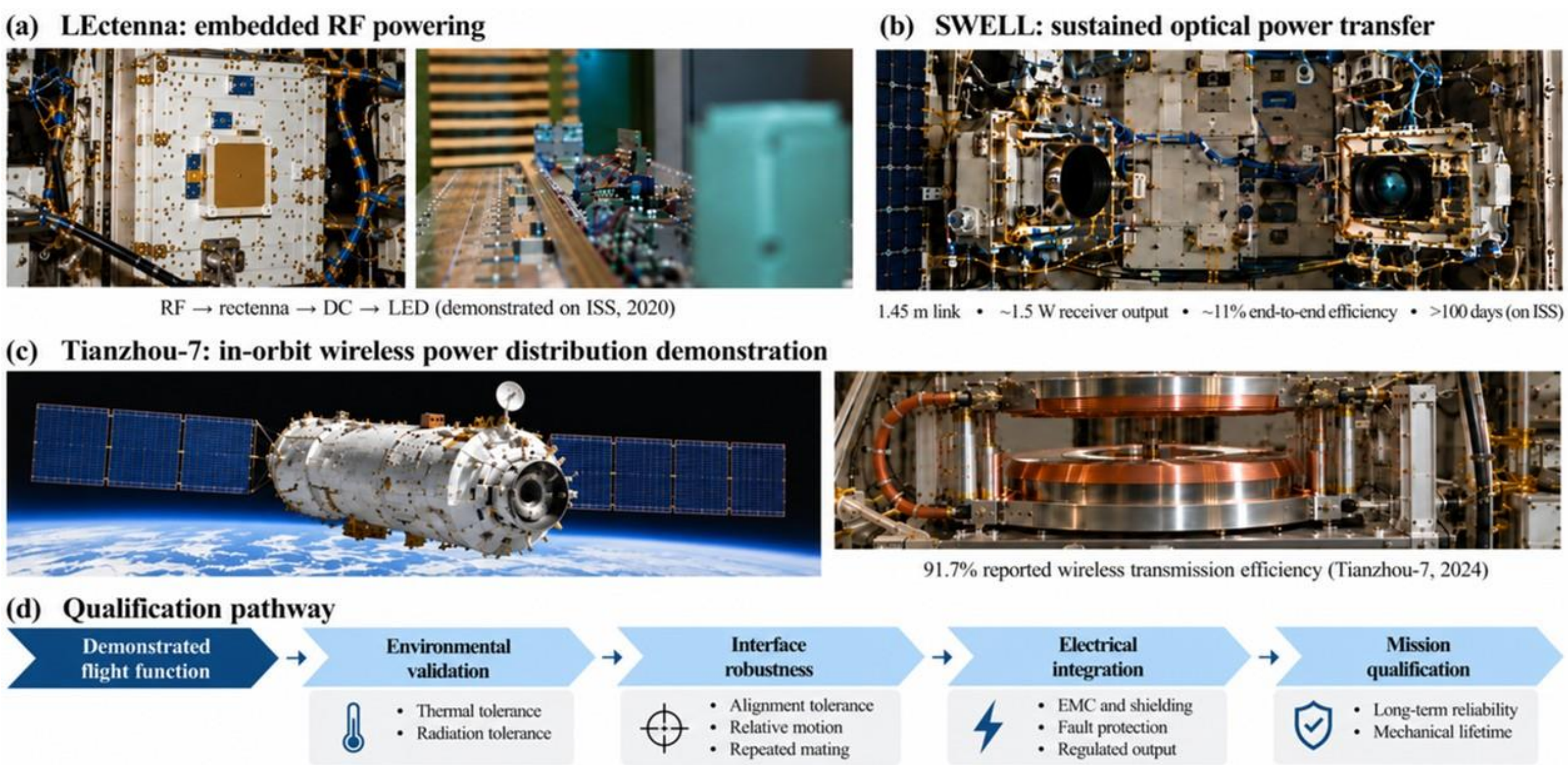

**Figure 13. Flight demonstrations and qualification pathway for intra-spacecraft wireless energy transfer.** (a) LEctenna demonstrated embedded RF powering aboard the ISS by converting incident microwave energy through a rectenna into DC power sufficient to operate an LED [114]. (b) The Space Wireless Energy Laser Link (SWELL) demonstrated sustained short-range optical power transfer over a 1.45 m link on the ISS STP-H9 platform, delivering approximately 1.5 W at the receiver output with a reported end-to-end efficiency of approximately 11% and more than 100 days of successful on-orbit operation [39]. (c) Tianzhou-7 demonstrated integrated near-field wireless power distribution in orbit, with a reported wireless transmission efficiency of 91.7% within the experimental boundary [40,41]. (d) Translation from demonstrated flight functions to operational spacecraft applications requires progressive environmental validation, interface robustness, electrical integration, and mission qualification. Reported efficiencies correspond to different physical mechanisms and measurement boundaries and are therefore not directly comparable.

These demonstrations establish different functions rather than directly comparable efficiency levels: LEctenna addresses embedded RF powering, SWELL sustained optical delivery, and Tianzhou-7 integrated near-field distribution. Figure 13 therefore compares demonstrated flight functions rather than ranking their efficiencies. As summarized in Fig. 13(d), these capabilities remain subject to mission-representative environmental, interface, electrical, and lifetime qualification, which is considered at the system level in Section 10.

In a summary, Figs. 12 and 13 distinguish interface-level feasibility from flight evidence. Figure 12 summarizes representative interface, receiver-boundary, and thermal-integration experiments, whereas Fig. 13 identifies complementary functions demonstrated in orbit and the remaining transition toward operational spacecraft use. Reliable conversion and power management remain necessary to close the pathway from demonstrated transfer to useful electrical service, as discussed in Section 9.

## 9. Key Enabling Technologies and System Integration

The previous sections establish the physical mechanisms, energy pathways, and mission-specific operating conditions of space WET. Translating these capabilities into an operational power system requires the integration of source conversion, beam control, receiver electronics, and downstream power management. Their performance is coupled through electrical, thermal, and environmental constraints, making useful-load power a more meaningful system metric than independently measured peak component efficiencies. This section examines the technologies governing that transition and distinguishes demonstrated hardware capabilities from the requirements of a space-qualified energy chain.

### 9.1 RF Source

The RF source converts electrical input into electromagnetic power. Amplifier performance is commonly characterized by drain efficiency or power-added efficiency (PAE), but neither metric accounts for all losses in the driving electronics, power distribution, combining networks, and antennas. Simons et al. [115] characterized a switched wideband K□-band GaN amplifier for future space-network applications. At 26.5 GHz, the device delivered 37.7 dBm saturated output power with 18.9% PAE; at 29.5 GHz, the corresponding values were 38.4 dBm and 23.3%. These measurements establish amplifier performance rather than the efficiency of a complete WET transmitter. Scaling to large radiating apertures introduces additional requirements for coherent power delivery. Structural deformation, thermal gradients, and attitude variations can alter the radiated field distribution and reduce receiver interception even when individual amplifiers maintain their nominal output. MAPLE demonstrated coherent operation of lightweight transmitting arrays in orbit [13], whereas JAXA/JSS investigated electronic phase compensation under displaced-array conditions [37]. Su et al. [116] subsequently developed predictive compensation for position and attitude deviations in ultra-large arrays and evaluated the approach using ground-based measurements without microwave power transmission. These studies address complementary aspects of transmitter development. Efficient RF generation must be maintained alongside aperture coherence and structural stability, requiring coordinated electrical, thermal, and control design at mission-relevant power and operating duration.

### 9.2 Rectennas and Waveform-Receiver Co-Design

AAt the RF receiver, useful DC output depends on incident power, impedance matching, waveform, and load. RF-to-DC efficiency must therefore be defined between the rectifier input and DC output terminals rather than treated as an interchangeable measure of rectenna or complete-link efficiency.

Yu et al. [117] investigated these operating dependencies using a 5.8 GHz GaN-on-SiC Schottky rectifier incorporating a harmonic-compression network. With a 400-Ω load, efficiency reached 73.3% at 40 dBm (10 W) and remained above 50% over approximately 22–42 dBm. Increasing the input to 43 dBm (20 W) raised the output voltage to 60.3 V but reduced efficiency to 45.5%. The experiment employed a heatsink and cooling fan; performance under spacecraft thermal and environmental conditions was not established. Such power-dependent behavior is particularly relevant to large rectenna arrays, where nonuniform illumination places individual elements at different operating points. Aggregate DC output consequently depends on element characteristics, impedance matching, and combining architecture. Frequency introduces further considerations. McSpadden et al. [56] reported 29% RF-to-DC efficiency at 35 GHz and 120 mW, whereas Yu et al. [117] investigated a substantially higher-power device at 5.8 GHz. Differences in device architecture and operating conditions prevent these results from isolating the effect of frequency. Waveform design provides another degree of freedom. Nonlinear energy-harvester modeling by Clerckx et al. [16] established a basis for joint waveform and receiver optimization. Wang et al. [118] subsequently demonstrated simultaneous wireless information and power transfer using a jointly modulated amplifying programmable metasurface, experimentally powering an LED array while transmitting video. Although conducted in a different power and application regime, this terrestrial experiment illustrates the value of coordinating transmitter modulation with receiver operation. Space rectennas must ultimately maintain useful DC output across the expected ranges of incident power, waveform, temperature, and load.

### 9.3 Optical Transmitters and Photovoltaic Receivers

Optical WET requires coordinated design of electrical-to-optical conversion, beam delivery, and photovoltaic reception. Beam quality, pointing stability, and illumination distribution must be matched to the receiver's spectral and thermal characteristics. Terrestrial experiments have advanced high-power optical transmission. Ochiai et al. [63] reported 1035 W of transmitted optical power over 1 km, producing 152 W of receiver electrical output and an approximately 15% efficiency under the authors' transmitted-optical-to-receiver-electrical definition. Beam shaping and receiver-side measures supported operation under daytime atmospheric turbulence. In a separate demonstration, DARPA's PRAD program [77] delivered more than 800 W over 8.6 km during a 30 s interval. These results establish specific high-power delivery capabilities, although neither constitutes qualification of a space laser system. Receiver studies reveal additional constraints. Ni et al. [119] reported 70.2% power conversion efficiency for a large-area six-junction GaAs laser power converter at −140 °C. A circular converter array achieved 50.6% laser-to-electrical efficiency at 25 °C, while a separate 50-m system reached a maximum electrical-to-electrical efficiency of 16.9%. These values correspond to different configurations and measurement boundaries and should not be compared directly. Temperature also affects spectral matching. Under cryogenic-vacuum conditions, Nouri et al. [120] measured efficiencies above 40% at 1064 nm over approximately 190–250 K. At 77 K, however, 980-nm illumination provided higher efficiency, consistent with the temperature-dependent widening of the semiconductor bandgap. Their thermomechanical analysis further identified material stress as a consideration under lunar thermal cycling. Laser wavelength and receiver architecture must therefore be selected together with the intended thermal operating range. Radiation introduces another dependence on device structure and operating history. Shi et al. [121] found that GaAs laser power converters with thicker bottom cells experienced greater electrical degradation following 1 MeV electron irradiation. Their simulations suggested that reducing bottom-cell thickness and optimizing its doping profile could improve radiation resistance. In single-junction InGaAsP converters, Aonuki et al. [122] observed an efficiency decrease from 39.8% to 22.8% after electron irradiation, followed by partial recovery to 28.7% after 30 min of 1064 nm illumination at 6.8 W/cm$^2$. Recovery was weaker at lower illumination intensity and negligible under the reported AM1.5G condition.

### 9.4 Power Management, Storage, and R6–R7 Closure

Electrical power emerging from an RF rectenna or photovoltaic receiver must be conditioned before reaching a spacecraft load. Voltage regulation, maximum power point tracking (MPPT), source switching, charge control, and fault protection introduce additional conversion stages and operating constraints. Power management and distribution (PMAD) therefore determine both the usable fraction of receiver output and the continuity of supply during link interruptions. Yigit et al. [123] demonstrated a 20-m terrestrial laser-power-transfer system incorporating aSiC LLC driver, optical transmission, a photovoltaic receiver, and MPPT. At the maximum-efficiency operating point, the driver reached 95.96% electrical efficiency, while 544.19 W of system electrical input produced 67.5 W of photovoltaic output, corresponding to 12.4% system efficiency. At a higher input of 875.08 W, photovoltaic output increased to approximately 98.8 W, but efficiency decreased to 11.29%. Thus, maximum output power and maximum system efficiency occurred at different operating points. The reported efficiency terminates at the photovoltaic output and does not independently quantify subsequent battery charging, discharge, regulation, or useful-load delivery. Although the experimental arrangement incorporated a battery unit, it did not establish a complete spacecraft-qualified R6–R7 pathway. Storage provides the temporal connection between intermittent wireless supply and load demand. Its required capacity depends on the received-power profile, interruption duration, charge–discharge losses, and allowable operating limits. Eclipse, orbital visibility, pointing interruptions, and network handovers may therefore impose different storage requirements even when average received power is similar. Closing the energy pathway requires measurement of regulated load energy over a defined operating interval, including PMAD and storage losses, source variability, and relevant fault conditions.

### 9.5 Hybrid Local Generation and Wireless Supply

Wireless delivery need not replace onboard generation and storage. In many mission concepts, WET is more appropriately integrated with local power sources: a rover may receive supplementary energy while retaining photovoltaics and batteries, a spacecraft may augment its onboard supply through beamed power, and distributed sensors may combine local harvesting with intermittent wireless recharge. Hybridization can improve service availability, but additional transmitting and receiving hardware, storage, thermal management, and control also increase system burden. Its value depends on whether the assembled architecture satisfies the required load schedule more effectively than alternatives providing the same electrical service. A high fraction of wirelessly supplied energy is not, by itself, evidence of a favorable design. The remaining task is to establish whether demonstrated component capabilities can be maintained when integrated under simultaneous mission constraints. This requires accounting for conversion losses, operating dependencies, source transitions, mass, and reliability across the complete energy pathway. Section 10 therefore addresses the qualification and system-level evidence needed to determine whether a WET architecture offers a measurable mission advantage.

## 10. Engineering Challenges and Future Perspectives

The remaining challenge in WET is to demonstrate reliable delivery of useful electrical energy under representative mission conditions. Five coupled issues govern this transition: (i) dominant losses; (ii) aperture and pointing scalability; (iii) environmental durability; (iv) operational availability, and (v) mission-level benefit.

The R0–R7 framework introduced in Section 3 provides a consistent basis for evaluating these issues. Its efficiency cascade is an accounting identity, not an assumption of independent stages, since temperature, loading, pointing, and control may affect several conversion processes simultaneously. High beam-collection efficiency has been demonstrated in the OMEGA prototype [75], receiver-side conversion has been investigated using GaN/Si rectifiers [112], and Yigit et al. [123] reported an integrated terrestrial laser-power-transfer system with an efficient driver. These results concern different energy boundaries and cannot be combined into a representative end-to-end efficiency. Moreover, the dominant loss varies with architecture, range, power level, and operating state. Optimization must therefore begin with a closed energy budget rather than an isolated component record. Aperture scaling introduces coupled electromagnetic and mechanical constraints. Larger microwave apertures improve directivity but increase deployment mass, feed-network complexity, structural deformation, and thermal sensitivity. Optical architectures can achieve narrow beams with smaller apertures, although they generally require tighter pointing and tracking tolerances. Su et al. [116] investigated predictive compensation for position and attitude deviations in large arrays, while MAPLE demonstrated coherent operation of a lightweight, flexible microwave array in orbit [13]. JAXA's ground experiments addressed precision microwave beam control [37]. Together, these studies establish complementary capabilities but do not resolve the combined effects of structural deformation, pointing errors, electrical losses, and receiver interception at full system scale. The relevant design objective is consequently reliable delivered power under representative disturbances, rather than nominal antenna gain or beam directivity alone. Thermal management must distinguish local dissipation from uncaptured radiative power. Conversion and distribution losses generate heat at their respective locations, whereas power missing the intended receiver may propagate away or be absorbed by other structures. Because heat rejection to the external environment in vacuum is primarily radiative, conversion performance is closely linked to radiator requirements and spacecraft configuration. Honda et al. [113] examined interactions between WET and spacecraft thermal insulation, whereas Nouri et al. [120] investigated photovoltaic conversion under cryogenic-vacuum conditions. Environmental evidence remains similarly fragmented: Tang et al. [49] studied gamma-irradiated GaN devices, Shi et al. [121]

examined electron-induced degradation in GaAs laser power converters, and Aonuki et al. [122] reported irradiation-dependent performance and partial recovery under laser illumination. Lunar-environment diffraction and scattering have also been modeled [95]. These studies identify distinct operating constraints but do not qualify a complete WET architecture under combined mission stresses. Future tests should reproduce mission-specific combinations of electrical loading, vacuum, radiation, thermal cycling, contamination, and structural disturbance, while measuring their effects on useful-load output. Operational endurance must likewise be distinguished from energy-service reliability. SWELL exceeded 100 days of successful on-orbit laser operation [39], while MAPLE was deployed for ten months, with experiments conducted during the latter eight months [13]. MAPLE demonstrated in-space power transfer and detection of an Earth-directed RF signal, but not useful electrical power delivery to a terrestrial load. These achievements provide important flight evidence without establishing continuous, multi-year operation of a complete WET energy chain. Reliability assessment should therefore extend beyond operating time to delivered energy and service availability over specified intervals, accounting for drift, misalignment, aging, interruptions, and fault recovery. Storage and power management must be included whenever they are required to maintain regulated load delivery. Existing demonstrations validate complementary portions of the energy pathway. Goldstone established high-power microwave reception and conversion [12]; JAXA investigated precision microwave beam control [37]; and PRAM demonstrated orbital solar-to-RF conversion without a remote power receiver [38]. SWELL advanced sustained in-space optical power transfer [39], MAPLE demonstrated flexible coherent microwave transmission, in-space power reception, and Earth-directed RF signal detection [13], and Tianzhou-7 extended near-field WET into integrated spacecraft power distribution [40, 41]. Figure 14 maps these complementary demonstrations onto the R0–R7 energy-reference planes, while Table 1 reports their measurement boundaries and evidence classes. Filled cells in Fig. 14 identify reference planes directly validated or reported within the experimental boundary of each study; unfilled cells indicate functions that were not directly validated within that boundary and should not be interpreted as inherent technical limitations. Their efficiencies are not directly comparable because the experiments address different operating conditions and portions of the energy chain. The principal validation gap is sustained source-to-load delivery under mission-representative environmental, pointing, thermal, storage, and load conditions.

Safety and electromagnetic compatibility impose further architectural constraints. Concentrated electromagnetic or optical fields must coexist with spacecraft communications, avionics, sensors, and scientific payloads. Vazquez Ramos [124] investigated radiated-emission compliance for spacecraft-oriented WET, while ITU-R SM.2505-1 [21] addresses RF-beam interference and human-hazard issues. Systems must control the intended beam together with sidelobes, harmonics, leakage, transients, and off-nominal emissions. ICNIRP [19] and IEEE C95.1-2019 [20] provide relevant exposure guidance, complemented by the mitigation approaches reviewed by Hirata et al. [18]. Nevertheless, mission-specific requirements concerning spectrum coordination, beam crossings, optical hazards, authorization, and fail-safe interruption require additional assessment. Safety and fault management should therefore be incorporated into architecture design and experimental validation rather than deferred to final certification.

Ultimately, WET must demonstrate a mission benefit beyond technical feasibility. Comparisons with local generation, storage, and wired distribution should use equivalent load demand, service duration, and reliability requirements while accounting for deployment mass, thermal control, auxiliary consumption, and operational complexity. Higher link efficiency alone is insufficient if these additional burdens outweigh the value of remote energy delivery. The decisive criterion is whether the integrated architecture provides the required useful energy within the constraints of its intended mission.

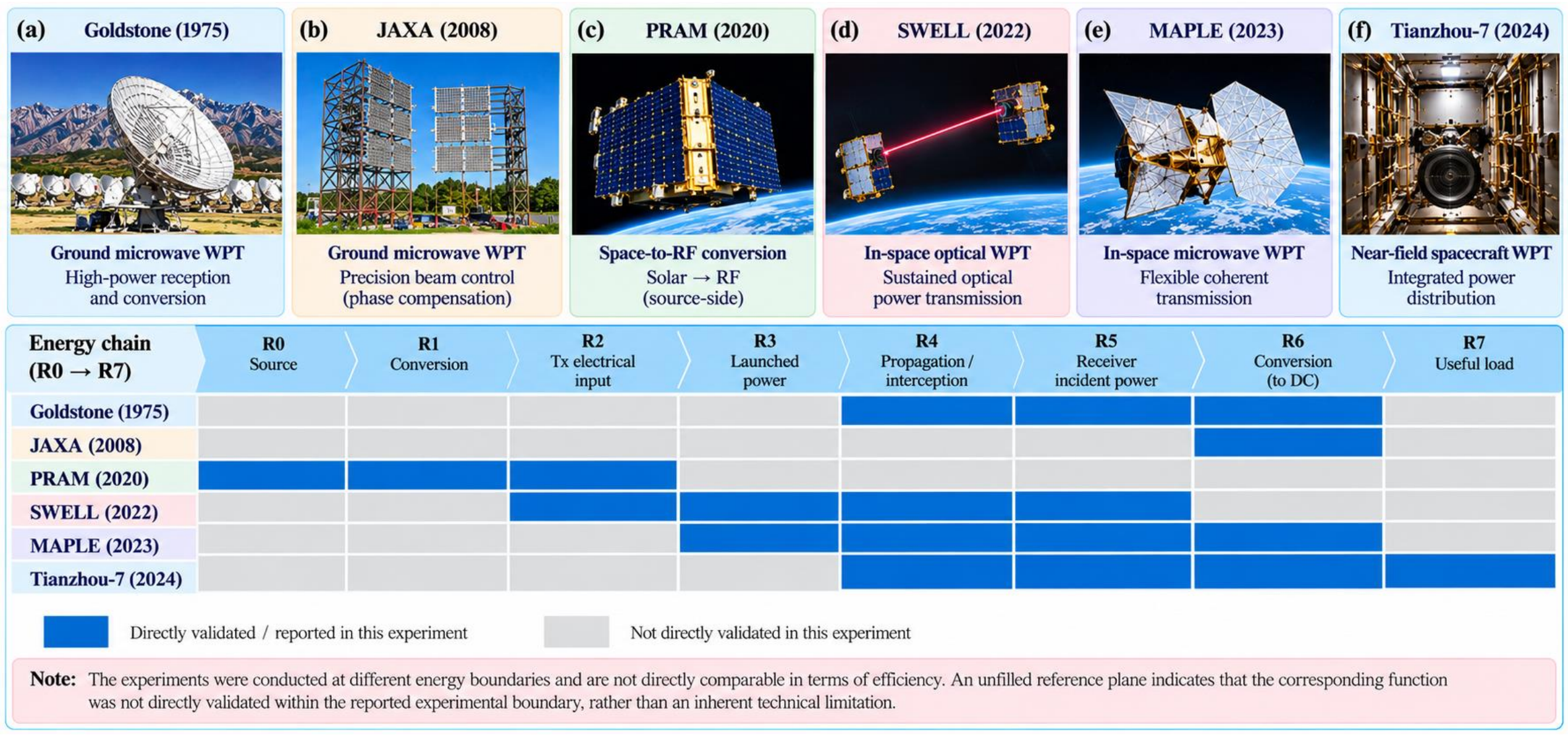

**Figure 14. Demonstrated functions and energy-boundary closure in representative WET experiments.** Representative ground and in-space demonstrations include (a) Goldstone [12], (b) JAXA [37], (c) PRAM [38], (d) SWELL [39], (e) MAPLE [13], and (f) Tianzhou-7 [40, 41]. The R0–R7 matrix maps the energy-reference planes directly validated or reported within the experimental boundary of each study. Filled cells denote directly validated or reported reference planes, whereas unfilled cells indicate functions not directly validated within the corresponding experimental boundary and should not be interpreted as inherent technical limitations. Reported efficiencies retain their original measurement boundaries and are therefore not directly comparable across experiments.

**Table 1. Quantitative performance and experimental maturity of representative space-WET systems. Reported values retain the original authors' measurement boundaries; evidence classes distinguish mission models, ground tests, and flight demonstrations.**

| Case | Carrier / range | Setting and quantitative result | Boundary / principal limitation | Evidence class | Ref. |
|---|---|---|---|---|---|
| **Kurs 2007** | Resonant magnetic; >2 m | 60 W; ~40% reported transfer efficiency | Authors' transfer-efficiency boundary; near-field coupling remains geometry-sensitive. | Ground experiment | [31] |
| **Goldstone** | 2.388 GHz; 1.54 km | 30.4 kW DC; peak incident 170 mW cm$^{-2}$; >80% receiver-array metric | Receiver-side metric; source-side conversion and R7 load not closed. | Integrated ground demo | [12] |
| **Sato 2025** | K-band; 150 mm | 14 W RF → 2.9 W DC; 20.7% | η(R1→R6); local 73.9% rectification does not represent the full link. | Ground experiment | [58] |
| **OMEGA** | 5.8 GHz; 55 m | 2081 W transmitted; 87.3% beam collection; 15.05% DC–DC | Beam collection and DC–DC are distinct; large loss remains outside spatial capture. | Integrated ground demo | [75] |
| **Mizrahi 2025** | 10 GHz microwave; GEO→Earth | Distributed close-flying GEO modules with dynamically programmable ground focal points; modeled LCOE 9.4 ¢/kWh after an assumed 10-year technology-development and scaling period | System/technoeconomic model; dispatchable delivery is projected rather than demonstrated as a closed orbital-to-ground electrical power chain. | System/technoeconomic model | [23] |
| **PRAD** | Optical; 8.6 km | >800 W delivered during 30 s; >1 MJ campaign transfer | Long terrestrial optical benchmark; no orbital pointing or spacecraft wall-plug closure. | Agency ground demo | [77] |
| **Ochiai 2025** | 1070 nm; 1 km | 1035 W optical → 152 W electrical; 15% | R1-type optical→R6 electrical; atmosphere/beam control remain integral to performance. | Integrated ground demo | [63] |
| **Xu 2024** | Laser; 10 m | 11.55% electric–electric efficiency | Authors' electric-to-electric metric; terrestrial 10-m test, not lunar environmental validation. | Ground experiment | [85] |
| **Washiro 2025** | Electric-field surface wave; sand-assisted near field | Prototype measurements using sand as a dielectric transmission medium; average transmission efficiency improved by about 6 dB | Ground prototype / lunar-surface analogue; not lunar-regolith or flight validation, and distinct from conventional plate-based CPT. | Ground prototype / conference study | [48] |
| **Turyshev 2026** | 1064 nm; cislunar→lunar surface | ~28 W daily-average receiver delivery in baseline case | Mission model; result depends on geometry, duty cycle, storage comparator, and client distribution. | System model | [84] |
| **RINGS** | Resonant inductive; ISS short range | ~30% preliminary on-orbit WPT | Preliminary on-orbit WPT metric; short-range resonant precursor, not radiative service. | Flight/microgravity precursor | [36] |
| **SWELL** | Optical; 1.45 m on ISS | ~1.5 W receiver output; ~11% reported end-to-end | Short-range flight link; ~1.5 W scale does not imply high-power bus replacement. | End-to-end flight link | [39] |
| **Tianzhou-7** | Near-field; in orbit | 91.7% reported wireless transmission efficiency | Reported transmission efficiency; exact electrical R-pair not publicly exposed. | Flight demonstration | [40, 41] |
| **MAPLE** | Microwave; LEO | Coherent WET/beamforming; Earth-directed signal detected | Flight beamforming subsystem; no useful-load closure. | Flight subsystem | [13] |
| **PRAM** | Solar→RF; X-37B | Source-side PV→RF sandwich-module experiment | Source-side flight constituent only; no remote receiver. | Flight constituent | [38] |
| **Li 2024** | 5.8 GHz metasurface router | ~3% best RF-to-RF PTE | RF-to-RF transmitter-integrated metric; not a remote relay. | Ground experiment | [8] |
| **Eidaks 2023** | RF active multi-hop relay | Received RF split between energy harvesting and RF amplification; harvested energy is conditioned/stored to support forwarding | Harvest-powered active relay; not a complete RF-to-DC-to-RF regeneration of the forwarded power; relay-side end-to-end efficiency not closed | Ground active-relay experiment | [10] |

## 10.1 Toward Distributed Space-Energy Infrastructure

The transition from individual WET links to distributed space-energy infrastructure is likely to proceed through incremental, hybrid architectures rather than an entirely wireless network. As illustrated in Fig. 15, a transmitter may initially supply a single load, subsequently serve multiple scheduled users, and ultimately participate in architectures incorporating relays, storage, coordinated energy management, and additional orbital or lunar nodes. This progression represents increasing architectural integration and coordination rather than a mandatory deployment sequence. Lunar power-system studies [87] and comparative assessments of surface power options [82] provide a basis for considering WET alongside local generation and storage. The latter stages in Fig. 15 are therefore presented as conceptual or prospective architectures rather than demonstrated operational infrastructure. Network expansion introduces requirements beyond those of isolated links. Shared sources must accommodate competing demands and interruptions, whereas routed delivery may introduce additional interception, conversion, storage, retransmission, and auxiliary-power losses. Passive redirection may avoid electrical conversion at the intermediate node, whereas regenerative relays require reception, electrical conversion or storage, and subsequent retransmission. Multi-node operation further requires scheduling, handover, state awareness, storage coordination, and fault management. Constellation-level power allocation [70] and lunar laser-beaming network optimization [93] have been investigated analytically, although their modeled performance does not establish operational network capability. Network assessment must therefore consider delivered load energy, service availability, storage requirements, conversion and routing losses, and infrastructure burden rather than link efficiency alone.

Multifunctional electromagnetic infrastructure offers a further direction. Li et al. [8] investigated reconfigurablemetasurface-based power routing, while the BEACON concept [94] considers the integration of beamed energy and communications. Wang et al. [118] demonstrated simultaneous wireless information and power transfer using a programmable amplifyingmetasurface under terrestrial conditions, and Zheng et al. [125] investigated reconfigurable holographic surfaces for space-oriented information and power transfer. These studies suggest opportunities to share apertures, tracking, scheduling, and waveforms across energy delivery, communication, and sensing. In prospective hybrid architectures, WET could operate together with local generation, storage, and other power sources rather than replacing them, while shared communication, navigation, or sensing functions could reduce duplicated infrastructure where mission requirements permit. Such integration is meaningful only when it improves measurable attributes, including mass, aperture utilization, auxiliary consumption, service availability, resilience, or useful delivered energy. Regardless of architectural complexity, the energy pathway should remain quantitatively traceable from source input to useful-load output. The R0–R7 framework can therefore be retained as the accounting basis for point-to-point links while being extended to identify additional conversion, storage, auxiliary-power, and routing boundaries introduced by shared-source, relay, networked, and hybrid architectures. The longer-term objective is not an entirely wireless power system, but an adaptable space-energy infrastructure in which WET complements local generation, storage, and conventional distribution while maintaining explicit energy accounting and mission-level benefit.

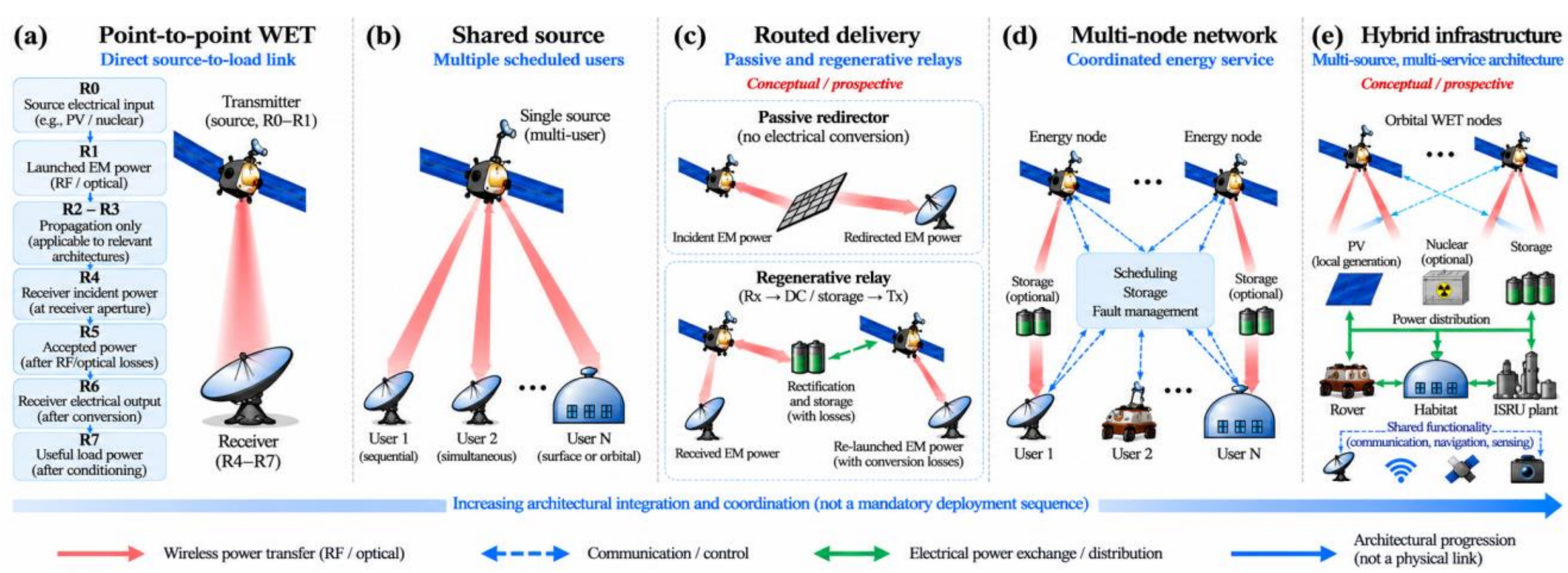

**Figure 15. Conceptual progression toward distributed space-energy infrastructure.** (a) Point-to-point WET with direct source-to-load energy accounting. (b) Shared-source service for multiple scheduled or simultaneously served users. (c) Routed delivery through passive redirectors or regenerative relays, the latter introducing additional conversion, storage, auxiliary-power, and retransmission requirements. (d) Coordinated multi-node operation incorporating storage, scheduling, communication and control, and fault management. (e) Prospective hybrid infrastructure integrating orbital or lunar WET with local generation, storage, conventional power distribution, and shared communication, navigation, or sensing functions. The progression indicates increasing architectural integration and coordination rather than a mandatory deployment sequence. Later stages are conceptual or prospective architectures rather than operationally demonstrated infrastructure. At constellation scale, orbital availability and ground-infrastructure constraints can further limit the number of realizable energy-delivery pathways, as illustrated by GEO SBSP capacity

modeling in Ref. [81]. Red solid arrows or beams denote wireless power transfer (RF or optical), blue dashed lines denote communication and control links, and green solid bidirectional arrows denote electrical power exchange or distribution. The blue solid horizontal arrow along the bottom denotes increasing architectural integration and coordination and does not represent a physical energy-transfer or communication link. Author-created synthesis based on the concepts and studies reviewed in the text.